\documentclass{article}

\usepackage{lipsum}
\usepackage{amsfonts}
\usepackage{amsmath,amssymb}
\usepackage{graphicx}
\usepackage{epstopdf}
\usepackage{algorithmic}
\usepackage[all]{xy}
\usepackage{caption}
\usepackage{url}
\usepackage{subcaption}
\ifpdf
  \DeclareGraphicsExtensions{.eps,.pdf,.png,.jpg}
\else
  \DeclareGraphicsExtensions{.eps}
\fi

\title{Computational modeling for high-fidelity coarsening of shallow water equations based on subgrid data\thanks{This work is part of the SPRESTO project, funded by the Dutch Science Foundation (NWO) in their TOP1 program}}

\author{Sagy R. Ephrati\footnote{Multiscale Modeling and Simulation, Faculty EEMCS, University of Twente, P.O. Box 217, 7500 AE Enschede, The Netherlands ({s.r.ephrati@utwente.nl})} 
\and Erwin Luesink\footnote{Department of Mathematics, Imperial College London, UK}
\and Golo Wimmer\footnote{Current affiliation: Los Alamos National Laboratory, New Mexico, USA}
\and Paolo Cifani\footnotemark[2]\hspace{1.5mm}\footnote{Gran Sasso Science Institute, Viale F. Crispi, 7 67100 L'Aquila, Italy}
\and Bernard J. Geurts\footnotemark[2]\hspace{1.5mm}\footnote{Multiscale Energy Physics, CCER, Faculty Applied Physics, Eindhoven University of Technology, P.O. Box 213, 5600 MB Eindhoven, The Netherlands} }

\begin{document}

\maketitle

\begin{abstract}
 Small-scale features of shallow water flow obtained from direct numerical simulation (DNS) with two different computational codes for the shallow water equations are gathered \textit{offline} and subsequently employed with the aim of constructing a reduced-order correction. This is used to facilitate high-fidelity \textit{online} flow predictions at much reduced costs on coarse meshes. The resolved small-scale features at high resolution represent subgrid properties for the coarse representation. Measurements of the subgrid dynamics are obtained as the difference between the evolution of a coarse grid solution and the corresponding DNS result. The measurements are sensitive to the particular numerical methods used for the simulation on coarse computational grids and can be used to approximately correct the associated discretization errors. The subgrid features are decomposed into empirical orthogonal functions (EOFs), after which a corresponding correction term is constructed. By increasing the number of EOFs in the approximation of the measured values the correction term can in principle be made arbitrarily accurate. Both computational methods investigated here show a significant decrease in the simulation error already when applying the correction based on the dominant EOFs only. The error reduction accounts for the particular discretization errors that incur and are hence specific to the particular simulation method that is adopted. This improvement is also observed for very coarse grids which may be used for computational model reduction in geophysical and turbulent flow problems.
\end{abstract}

\paragraph*{Keywords}
  empirical orthogonal functions,
  data-driven subgrid modeling,
  numerical simulation, finite volume discretization, finite element discretization, shallow water equations, multiscale modeling

\paragraph*{AMS}
  76B99, 65M22, 86-10

\section{Introduction}
The nonlinear nature of models in fluid dynamics causes small scale and large scale flow features to interact with each other. This implies that one would need to resolve the entire range of scales from the largest down to the smallest dynamically relevant Kolmogorov scale present in the particular problem, in order to have a good fluid-mechanical model. In geophysical fluid dynamics, typical largest length scales are in the order of hundreds of kilometres. This means that solving the entire range of scales down to the Kolmogorov length scale is by far too expensive for modern-day high performance computing. Any feasible approach will hence necessarily have to imply simplifications, either in the completeness of the mathematical model or in the spatial and temporal resolution with which the dynamics is approximated, or both. In this paper we will work out an offline/online approach in which we use explicit knowledge of the smallest scale dynamics obtained from prior \textit{offline} fully resolved simulations, in order to arrive at an \textit{online} computational high-fidelity coarsening. This approach is illustrated for the shallow water equations in which we opt for an empirical orthogonal function (EOF) representation of the corresponding subgrid forcing. The accuracy and efficiency we find for this approach and the rate with which the EOF representation converges in selected cases, establishes the feasibility of this computational model reduction for shallow water models.

There is a strong interest into the coarsening of detailed computational models in order to reach predictions and simulations that are on the one hand of sufficient accuracy for a particular problem, while requiring considerably less effort in terms of time and storage compared to the underlying detailed description \cite{geurts2003elements}. These problems are at the core of the field of `Reduced Order Modeling' (ROM) \cite{burkardt2006pod}. A prominent example is so-called large-eddy simulation (LES) in  which the spatially filtered Navier-Stokes equations form the point of departure for large-scale models that can handle turbulent flow at high Reynolds numbers \cite{sagaut2006large}. The filtering of the nonlinear terms in the Navier-Stokes equations introduces a closure problem and additional high-pass smoothing associated with the spatial discretization method \cite{geurts_vdbos}. These aspects are typically addressed by the introduction of a subgrid scale model to represent the influence of the smaller scale dynamics on the retained resolved scales. The design of good subgrid parameterizations is challenging and LES models based on physical arguments are often based on a crude approximation of the actual subgrid dynamics. Moreover, artificial dissipation introduced by the truncation error of the coarse-grid PDE may be dominant leading to an over-dissipative system. 

In this paper, we approach the problem of achieving accurate and effective coarsened flow models differently. Here, by introducing an explicit subgrid scale forcing extracted from a previously conducted direct numerical simulation (DNS) of the same problem, we account for the accumulated effects of the unresolved dynamics. Using high-resolution data to find subgrid parametrizations has been applied to, e.g., oceanic flows \cite{bolton2019applications} and atmospheric processes \cite{o2018using}. By adding a corresponding correction term to the governing equations, an alternative representation of the small scale dynamics is obtained. This paper is strongly motivated by the seminal work that led to the so called stochastic advection by Lie transport (SALT) approach and pursues the path of introducing tailored forcing to the equations in order to account for missing dynamics in the coarsened solution.  In \cite{holm2015variational} a stochastic variational principle was introduced to derive equations in continuum mechanics in such a way that the geometric structure corresponding to these equations remains the same. 
The SALT method has important applications in geophysical fluid dynamics, for instance to address the fundamental problem of appropriately representing measurement error and uncertainty due to neglected physical effects, spatial and temporal coarsening of the dynamics, and incompleteness of the mathematical model. In \cite{holm2015variational} the subgrid dynamics are computed from the difference between fully resolved and filtered Lagrangian trajectories. Here we construct a coarse-grid correction from the difference between the solution of the fine PDE and the coarse PDE at given time instances. The latter allows to take into account not only the effect of the subgrid scales but also the numerical error. 

Analogously to \cite{holm2015variational}, in this work we represent the coarse-grid correction by means of empirical orthogonal function (EOF) analysis \cite{lumley1967}. The subgrid term structure is thus captured by the solution eigenvectors to the EOF problem, henceforth called $\boldsymbol \xi_i$. Differently from \cite{holm2015variational}, no stochasticity is introduced here into the model and the effect of the coarsening is modelled as a deterministic forcing.

The technique of EOF analysis is well-known in atmospheric and oceanic dynamics, and is often called proper orthogonal decomposition (POD) in the context of fluid dynamics \cite{tennekeslumley}. EOF analysis has been applied in atmospheric sciences since the 1950s, for instance in  \cite{hannachi2007empirical}, \cite{lorenz1956empirical}, with the purpose of identifying coherent structures in the solution and reducing dimensionality of weather and climate systems. Examples of applications in fluid dynamics include the analysis of canonical problems in turbulence such as the lid-driven cavity \cite{cazemier98}, the turbulent jet \cite{meyer2007} and channel flow \cite{muralidhar2019spatio}. Instead of using the EOF method to analyze flow structures, we apply it to construct a basis for the coarse-grid correction.  
We illustrate the method with shallow water flow under the influence of external agitation, complementing the earlier work on the Euler equations in periodic domains \cite{cotter2019numerically}.

By construction the coarse-grid correction is dependent on the adopted numerical method. Hence, we will investigate two different methods for solving the shallow water equations and compare the type and size of EOF corrections needed to improve a coarse simulation. Moreover, the convergence of the corrections upon increasing the number of EOFs will be investigated. In the SALT approach, one investigates differences only in the velocity variables, since one introduces stochasticity in the vector fields that carry the flow properties. 
Results of \cite{holm2019stochastic} imply that for this situation, obtaining the $\boldsymbol \xi_i$ in one dimension and extending their domain to two dimensions corresponds to $\boldsymbol \xi_i$ obtained from the two-dimensional translation-invariant setting.

The following is an overview of the key results discussed in this paper: \begin{itemize}
    \item A subgrid data measurement procedure is presented, applicable to any set of PDEs, here applied to the shallow water equations. These measurements are extracted from an \textit{offline} computation of the fine and coarse PDE.
    \item Subgrid data is measured for two test cases which are both performed using a finite difference discretization and a finite element discretization. The test cases feature a submerged ridge as bathymetry and include constant external forcing (first test case) and periodic external forcing (second test case). The subgrid data are decomposed into EOFs and their corresponding time series.
    \item The level of approximation of the original dataset when applying different numbers of EOFs is investigated for the test case with external forcing.  A coarse numerical solution with zero error is obtained when the full set of EOFs is used. Truncating the reconstructed correction term to a subset of the EOFs significantly reduces the error on coarse computational grids, independent of the used numerical method.
    \item A significant error reduction is obtained when applying the developed reduced-order correction method to the same test case with different initial conditions. This indicates that the measured temporal coefficients tolerate some level of approximation without significant loss of accuracy.
\end{itemize}
The paper is organised as follows. In section \ref{sec:governingequations}, we will introduce the governing equations as well as the discretisation methods that will be used to simulate the governing equations. 
Section \ref{sec:measurements} describes the measuring procedure and the reduced-order model. In section \ref{sec:eofs} we investigate the convergence of the EOF decomposition of the coarse-grid correction for two test cases: a steady flow (subsection \ref{subsec:testcase1}) and a periodically forced flow (subsection \ref{subsec:testcase2}) over a bathymetry represented by a Gaussian profile. In section \ref{sec:rom} the developed reduced-order model is applied to the test cases of section \ref{sec:eofs}. In particular, a range of grid resolutions is investigated as well as the behavior of the model for a varying number of EOFs. Finally, the measured corrections are applied to the same problem with perturbed initial conditions (subsection \ref{subsec:ROM_ICs}) and accuracy in the prediction of long-time averages is investigated (subsection \ref{subsec:ROM_statsteadystate}).
In section \ref{sec:conclusion} we conclude the paper and formulate future challenges in the outlook.

\section{Governing equations and numerical methods}
\label{sec:governingequations}
The model that is central to this work is the shallow water (SW) model. The SW equations, also called the Saint-Venant equations, describe the behaviour of a fluid in a shallow channel with a free surface and bottom topography. This model can be derived by vertically integrating the incompressible free surface Euler-Boussinesq equations over the shallow domain in the small aspect ratio limit, as is demonstrated in \cite{holm2019stochastic}. The SW model is nonlinear and consists of two coupled equations. The first equation describes the evolution of the velocity $u$ and the second equation is the continuity equation that describes the evolution of the water depth $\eta$. The total depth is the difference between the free surface elevation $\zeta$ and the bottom topography (or bathymetry) $b$, hence $\zeta=\eta-b$. Additionally, we will consider external forcing and damping of the velocity. In one spatial dimension the SW model with forcing and damping is given by
\begin{equation}
\begin{aligned}
u_t + \frac{1}{2}(uu)_x + \frac{1}{\mathrm{Fr}^2}(\eta - b)_x & = a(t)-ru, \\
\eta_t + (u\eta)_x & = 0.
\end{aligned}
\label{eq:rsw}
\end{equation}
The right-hand side of the momentum equation contains a time-dependent forcing term $a(t)$ and a damping coefficient $r$ which induces damping proportional to the velocity. Here ${\rm Fr}$ is the Froude number, which is defined as the ratio between the typical velocity scale $U$ and the fastest gravity wave $\sqrt{g H}$, where $H$ is the typical depth and $g$ is the gravitational acceleration. For the study of this paper the one-dimensional model is a suitable formulation, combining low computational cost with a truthful representation of the underlying dynamics. In fact, this model is directly related to the two-dimensional rotating shallow water equations, which form a convenient model in geophysical fluid dynamics. It is known as the simplest model that incorporates the interaction between Rossby waves and gravity waves at geostrophic balance \cite{zeitlin2018geophysical}.

In the following we provide a description of the two numerical methods that are used in this study. The two corresponding methods are based on finite difference (FD) and a finite element (FE) discretization methods used for solving nonlinear PDEs and are employed here (i) to investigate convergence of the obtained numerical solutions and (ii) subgrid measurements, and (iii) to show the application of reduced-order corrections. 

The main difference between the methods is that the FD method solves the momentum equation with first-order accuracy and the continuity equation with second-order accuracy, whereas the FE method solves these equations with second and first-order accuracy, respectively. The main benefit of the FD method is its simplicity and low computational cost, while the FE method is easily extendable to problems in more dimensions and on complex geometry. The approach demonstrated in this paper is general and extendable to different numerical methods other than those analysed here.

The time integration is the same for both discretizations and is given by a fourth order Runge-Kutta method (RK4). 
The time-step is specified to satisfy numerical stability, which yields temporal discretization errors that are considerably smaller than the spatial discretization errors.

\subsection{Collocated finite difference discretization (FD)}
The finite difference discretization is based on a collocated arrangements of the discrete variables $(u_i,\eta_i)$ approximating the exact solution $(u(x_i),\eta(x_i))$ at the grid nodes $x_i$ with $i$ running from 0 to $N$, corresponding to an Arakawa A-grid \cite{arakawa1977computational}. The first-order upwind method has been employed for the discretization of convection of momentum. This provides numerical stability of the resulting discrete hyperbolic partial differential equation. The pressure term and the continuity equation are discretized using second-order central differences. Conservation of mass is ensured by discretizing the conservative form of the continuity equation.
The finite difference discretization is summarized as  \begin{equation}
    \begin{aligned}
        \frac{1}{2}(uu)_x\big|_{x_i} = (u u_x)_{x_i}&\approx \begin{cases} u_i\left(u_i - u_{i-1}\right)/\Delta x \hspace{3mm}\text{ if $u_i>0$,} \\
        u_i\left(u_{i+1}-u_i \right)/\Delta x \hspace{3mm}\text{ if $u_i<0$,}
        \end{cases} \\
        \left(\eta - b\right)_x\!\big|_{x_i} & \approx \left(\eta_{i+1} - b_{i+1} - \eta_{i-1} + b_{i-1} \right) / \left(2 \Delta x\right), \\
        \left( u\eta\right)_x\!\big|_{x_i} & \approx  \left(u_{i+1}\eta_{i+1} - u_{i-1}\eta_{i-1} \right)/\left(2\Delta x \right),
    \end{aligned}
    \label{eq:FDmethod}
\end{equation}
with $\Delta x$ the grid size. No modification  of the numerical scheme \eqref{eq:FDmethod} is required at the boundary, since periodic conditions are imposed.
The discretized momentum equation has a formal order of accuracy of one, due to the chosen discretization of the convective term. The continuity equation is second-order accurate. 

\subsection{Compatible finite element discretization (FE)} 
The finite element discretization is given by a mixed compatible element method, which can be seen as a finite element version of a finite difference discretization based on an Arakawa C grid \cite{arakawa1977computational}. It has been proposed as a discretization method for numerical weather prediction in \cite{COTTER20127076, natale2016compatible}, as it inherits the desirable properties of the C-grid -- such as exact steady geostrophically balanced states for the linearized shallow water equations. A description of the method can be found in appendix \ref{Appendix:FEM}.

A pair of compatible spaces for $u$ and $\eta$ is given, e.g., by
\begin{equation}
    \mathbb{V}_u = CG_k(\Omega), \hspace{1cm} \mathbb{V}_\eta = DG_{k-1}(\Omega),
\end{equation}
where $CG_k(\Omega)$ denotes the $k^{th}$ polynomial order continuous Galerkin space and $DG_{k-1}(\Omega)$ the $(k-1)^{th}$ polynomial order discontinuous Galerkin space.

The governing shallow water equations \eqref{eq:rsw} are discretized such that the divergence in the continuity equation is considered strongly, while the gradient in the momentum equation is imposed weakly, leading to the mixed formulation
\begingroup
\addtolength{\jot}{2mm}
\begin{align}
    &\left\langle w, u_t \right\rangle - \left\langle w_x, \frac{1}{2}u^2 + \frac{1}{\text{Fr}^2}(\eta - b) \right\rangle = 0 & \forall w \in \mathbb{V}_u, \label{doc_cfem_u_eqn_no_upw}\\
    & \eta_t + F_x = 0, \label{doc_cfem_eta_eqn_no_upw}
\end{align}
\endgroup
where $\langle . ,. \rangle$ denotes the $L^2$ inner product, and the flux $F$ in~(\ref{doc cfem_eta_eqn_no_upw}) is given by the $L^2$-projection of $\eta u$ into the velocity space, i.e.,
\begin{align}
    \langle w, F - \eta u \rangle = 0 && \forall w \in \mathbb{V}_u.
\end{align}

The above space discretization conserves mass locally as well as a discrete energy globally (for details, see e.g. \cite{mcrae2014energy}). In this paper, we consider the lowest polynomial order $k=1$ for this setup.
\section{Data measurements and processing}\label{sec:measurements}
This section describes the procedure of measuring the subgrid data and subsequently constructing a reduced-order correction based on these measurements. Given a truth $u_\mathrm{truth}$ and a coarse-grid result $u_\mathrm{sim}$, we construct a function $f(x,t)$ via
\begin{equation}
    u_\mathrm{truth}(x,t) - u_\mathrm{sim}(x,t) = f(x,t) = \bar{f}(x) + f'(x,t)
    \label{eq:utruth-usim}
\end{equation}
where the measurements are decomposed into a mean $\bar{f}(x)$, which will be referred to as $\boldsymbol \xi_0(x)$, and a fluctuating component $f'(x,t)$. The EOF decomposition is applied to the fluctuating component $f'$, which is assumed to be stationary in the average or statistical sense. Specifically, on a numerical grid consisting of $N$ cells, this algorithm yields $N$ eigenmodes $\boldsymbol \xi_i(x)$ with corresponding temporal coefficients $\alpha_i(t)$: \begin{equation}
    f'(x,t) = \sum_{i=1}^N \alpha_i(t)\boldsymbol \xi_i(x).
    \label{eq:EOF_decomp}
\end{equation}
The measuring procedure described below is such that it identifies the features missing from a (coarse) numerical solution. The constructed $f(x,t)$ can be introduced into coarse simulations as a forcing or correction term, thus correcting the numerical solution to match the reference truth. In the ideal setting, all data is available and the numerical solution can be corrected so that it perfectly recovers the truth on the coarse grid. However, this is typically not feasible in practice due to large data storage requirements. The EOF approach allows for an optimal approximation of the entire data set using a finite number of modes.

This section presents this methodology as follows. The subgrid term measuring procedure is given in  \ref{subsec:measurement_procedure} and section \ref{subsec:EOF} briefly summarizes the EOF algorithm. Subsequently, the reduced-order correction is detailed in section \ref{subsec:DefiningROM}.

\subsection{Subgrid term measurement procedure}
\label{subsec:measurement_procedure}
A simulation, which will correspond to a dataset, runs from time $t=0$ to $t=T$. The measuring intervals are indicated by $\Delta t_M$ and are such that $N_M\Delta t_M = T$, where $N_M$ denotes the number of measuring intervals. For consistency, the coarse-grid time step $\Delta t$ is set to be equal to $\Delta t_M$. The measurements comprise of the difference of the evolution of the true velocity and free surface height $(u_\mathrm{truth},\eta_\mathrm{truth})$ and their corresponding coarse-grid numerical solution $(u_\mathrm{sim},\eta_\mathrm{sim})$, as in equation \eqref{eq:utruth-usim}. The truth is calculated by performing a numerical simulation on a very fine grid. Throughout this study a grid consisting of 512 computational cells is considered sufficiently fine to accurately resolve all scales of motion. This has been verified by conducting a grid refinement study.

The numerical coarse grid solution $(u_\mathrm{sim},\eta_\mathrm{sim})$ is the quantity that we wish to improve. Since the coarse grid solution and the truth are defined on different computational grids, comparing the two solutions is done by restricting $(u_\mathrm{truth},\eta_\mathrm{truth})$ to the grid on which $(u_\mathrm{sim },\eta_\mathrm{sim})$ is defined. This is carried out by introducing a restriction operator $R$, here chosen to be equal to the injection of fine-grid values onto coarse-grid values.

The subgrid term defined for the velocity and free surface height will be denoted by $\mathbf{f}(x,t)=(f_u(x,t), f_\eta(x,t))$.
Let us assume $u_\mathrm{truth}$ at time $t_0$ to be known. The subgrid correction over a time-interval $[t_0,t_0+\Delta t_M]$ is estimated by applying the following procedure.
\begin{enumerate}
    \item Inject the truth to the coarse grid at $t=t_0$ and set $u_\mathrm{sim}(x,t_0)=Ru_\mathrm{truth}(x,t_0)$ and $\eta_\mathrm{sim}(x,t_0)=R\eta_\mathrm{truth}(x,t_0)$, with $R$ a coarse-graining operator.
    \item Integrate the fine and coarse grid solution from $t=t_0$ to $t=t_0+\Delta t_M$. 
    \item Evaluate
    \begin{align}
\nonumber    f_u(x,t_0+\Delta t_M) &= Ru_\mathrm{truth}(x,t_0+\Delta t_M)-u_\mathrm{sim}(x,t_0+\Delta t_M) \\ 
&=R\left(\int_{t_0}^{t_0+\Delta t_M}\! u_{t,\mathrm{truth}}\,\text{d}t\right) - \int_{t_0}^{t_0+\Delta t_M}\! u_{t,\mathrm{sim}} \,\text{d}t, \label{eq:definition_f}
\end{align}
\end{enumerate}
and analogously for $f_\eta(x,t_0+\Delta t_M)$. These measurements are done offline. In the next subsections we describe how the measurements are processed and subsequently applied \textit{online} as a correction term in coarse numerical simulations.

\subsection{Empirical Orthogonal Function Analysis}
\label{subsec:EOF}
The measurements $\mathbf{f}$ are stored in a matrix $\mathbf{V}^N\in \mathbb{R}^{M\times N}$, where $M$ is the number of coarse grid points and $N$ is the number of measurements. The entry $(\mathbf{V}^N)_{ij}$ corresponds to the subgrid difference at grid point $x_i$ at the $j^\mathrm{th}$ measuring instant. 
The time-mean from $M$ time series is subtracted from the matrix $(\mathbf{V}^N) \in \mathbb{R}^{M\times N}$ to form the anomaly matrix $\mathbf{A}$, whose rows have zero mean. The time-mean is the spatial profile previously introduced as $\boldsymbol \xi_0$. One would then compute the covariance matrix $\mathbf{R} = \mathbf{A}\mathbf{A}^T$ and solve the eigenvalue problem 
\begin{equation}
\mathbf{R}\mathbf{C} =\mathbf{C}\boldsymbol{\Lambda},
\label{eq:eigR}
\end{equation}
where the columns of $\mathbf{C}$ are the eigenvectors $\boldsymbol\xi_i$ (EOFs) and the eigenvalues (EOF variances) are on the diagonal of $\boldsymbol \Lambda$. 
A drawback of this method is that computing the covariance matrix becomes very numerically expensive as the amount of stored data rapidly increases with the number of snapshots. This can be dealt with by computing the SVD of the anomaly matrix \cite{dawson2016eofs, trefethen1997numerical}. Subtituting $\mathbf{A}=U\Sigma V^T$ into the the definition of the covariance matrix yields \begin{equation}
\mathbf{R} = \mathbf{A}\mathbf{A}^T =U\Sigma\Sigma^T U^T. \label{eq:R_AAT_SVD}
\end{equation}
Comparing equations \eqref{eq:eigR} and \eqref{eq:R_AAT_SVD}, it is observed that $\mathbf{C}=U$ and $\boldsymbol \Lambda = \Sigma\Sigma^T$. 

An insufficient number of measurements leads to statistical error in the computation of the covariance matrix. In this study, it is assumed a sufficient number of measurements is available for the EOF algorithm.

\subsection{Defining a reduced-order correction for the SWE}
\label{subsec:DefiningROM}
Having introduced the measurement procedure and the EOF algorithm, we can now define a correction term based on the decomposed measurements. This term is included in the numerical simulation such that, if all available data is used, the corrected coarse solution would equal the truth on the coarse grid. We denote the EOFs for the velocity and free surface height by $\xi_{i,u}$ and $\xi_{i,\eta}$, respectively, with corresponding time series $\alpha_{i,u}$ and $\alpha_{i,\eta}$. The correction based on $n$ EOFs is denoted by $\left(f_{n,u}(x,t), f_{n,\eta}(x,t) \right)$ for $u$ and $\eta$ individually, where
\begin{equation}\begin{aligned}
    f_{n,u}(x,t) &= \xi_{0,u}(x,t) + \sum_{i=1}^n\alpha_{i,u}(t)\xi_{i,u}(x), \\
    f_{n,\eta}(x,t) &= \xi_{0,\eta}(x,t) + \sum_{i=1}^n \alpha_{i,\eta}(t)\xi_{i,\eta}(x).
    \end{aligned}
    \label{eq:correctionterm}
\end{equation}
For an explicit Euler scheme, the reduced-order model is formulated as follows:
\begin{equation}
    \begin{aligned}
    u^{k+1} &= u^k + \Delta t \mathcal{L}(u^k,\eta^k) + f_{n,u}^{k+1}, \\
    \eta^{k+1} &= \eta^k + \Delta t\mathcal{D}(u^k,\eta^k)+f_{n,\eta}^{k+1},
    \end{aligned}
    \label{eq:reducedorderequations}
\end{equation}
where $k$ is the time level, $\mathcal{L}$ is the discrete differential operator of $-\frac{1}{2}(uu)_x - \frac{1}{\mathrm{Fr}^2}(\eta-b)_x + a(t)-ru$, $\mathcal{D}$ is the discrete divergence $(u\eta)_x$ and $f_n^{k+1}$ is the correction measured at time $k+1$ over an interval $\Delta t$ and decomposed into $n$ EOFs. Extension of \eqref{eq:reducedorderequations} to RK4 is straightforward. 

Finally, the temporal coefficients are obtained by projecting the governing equations on the spatial structures. Given an inner product $\langle \cdot, \cdot \rangle$, $\alpha_i(t)$ can be determined from $\langle f'(x,t), \xi_i(x)\rangle$ when the decomposition \eqref{eq:EOF_decomp} is used. 
In matrix notation, this is given by
\begin{equation}
    \boldsymbol\alpha = \mathbf{A} \mathbf{C}. \label{eq:temporalcoefficients_matrix}
\end{equation}

The algorithm for computing and applying the subgrid corrections is summarized as follows: \begin{enumerate}
\item The difference between the reference (fine-grid) evolution and coarse-grid evolution are measured, as per \eqref{eq:definition_f}.
\item The measurements are stored in a matrix which serves as input for the EOF algorithm.
\item An $n^\mathrm{th}$-order correction term is constructed by considering the time-mean and the first $n$ EOFs, by means of \eqref{eq:correctionterm}.
\item The corrections are applied to the coarse numerical solution after completing a time step, as in \eqref{eq:reducedorderequations}.
\end{enumerate}

\section{Convergence analysis of EOFs of subgrid data}\label{sec:eofs}
In this section, we present the results of simulations using the two numerical methods for the shallow water equations introduced in Section~\ref{sec:governingequations}. A comparison is performed for two test cases for which the subgrid corrections on several coarse grids are determined. The bathymetry for both test cases is defined by $b(x)=1-A\exp\left(\frac{-(x-0.5L_x)^2}{B^2}\right)$. The latter describes a submerged ridge of height $A$ and width $B$. The values for $A$ and $B$ are $0.01$ and $0.15$, respectively. The initial conditions are $u(x,0)=0$ and $\eta(x,0)=b(x)$. 

We force the flow differently in both tests. The first case uses a constant forcing, modeling a fixed `tilting' of the entire domain. Damping is added to keep the flow bounded. 
For the second case a time-periodic external forcing is applied to emulate tidal behaviour or `sloshing'. 
The Froude number for each test case is fixed at $\mathrm{Fr}=0.75$ to steer away from the possibility of shocks occurring in the solution. The latter behaviour is not the focus of this paper. 

In the analysis of the results, all $\boldsymbol \xi_i$ are multiplied by the square root of the corresponding eigenvalues and convergence of the $\boldsymbol \xi_i$ is quantified by comparing the infinity norm of the eigenfunctions on various grids.

The reference solution is defined as the numerical solution on a grid of 512 computational cells. 
The corresponding coarse simulations range from 256 down to 8 grid cells. The ratio between the coarse and fine time steps size is fixed at 4. 
For all simulated coarse grids one could choose a different $\Delta t$ on each grid to ensure stability. Since the method used here is general and applies for any value of $\Delta t$, for convenience and without loss of generality we have adopted the same time step size for all grids. 

\subsection{Steady flow over a periodic ridge}
\label{subsec:testcase1}
The steady flow over a periodic ridge can be computed reliably at a range of spatial resolutions, using both simulation methods. Here we analyze the profiles of the eigenvectors $\boldsymbol\xi_i$ and the energy associated to them for different grid coarsenings.

By introducing forcing and a counterbalance damping, which emulates tilting of the domain, the model reaches a nontrivial stationary state. 
In a practical setting, the damping can be thought of as a necessary term to control the discharge rate of the fluid. From this point the measurements of the coarse-grid correction are gathered. For a value of the forcing and damping rates ($a$ and $r$ in equation \eqref{eq:rsw}) equal to $0.5$, an approximately steady state is reached at $t=30$. Measurements are then collected for one time unit, a time interval deemed sufficiently long to generate enough data for the EOF algorithm.  

Since the flow is at steady state, the time mean $\boldsymbol\xi_0$ in equation \eqref{eq:correctionterm} captures virtually all of the coarse-grid difference that should be added, at each time-step, to maintain the fine solution on the coarse grid. Ideally only the coarse-grid correction after one time-step is needed to recover a steady solution. However, given the fact that the fine grid solution is still varying slowly, we accumulate measurements over one time unit.

\begin{figure}[h!]
    \centering
    \begin{subfigure}{0.45\textwidth}
        \centering
        \includegraphics[width=\textwidth]{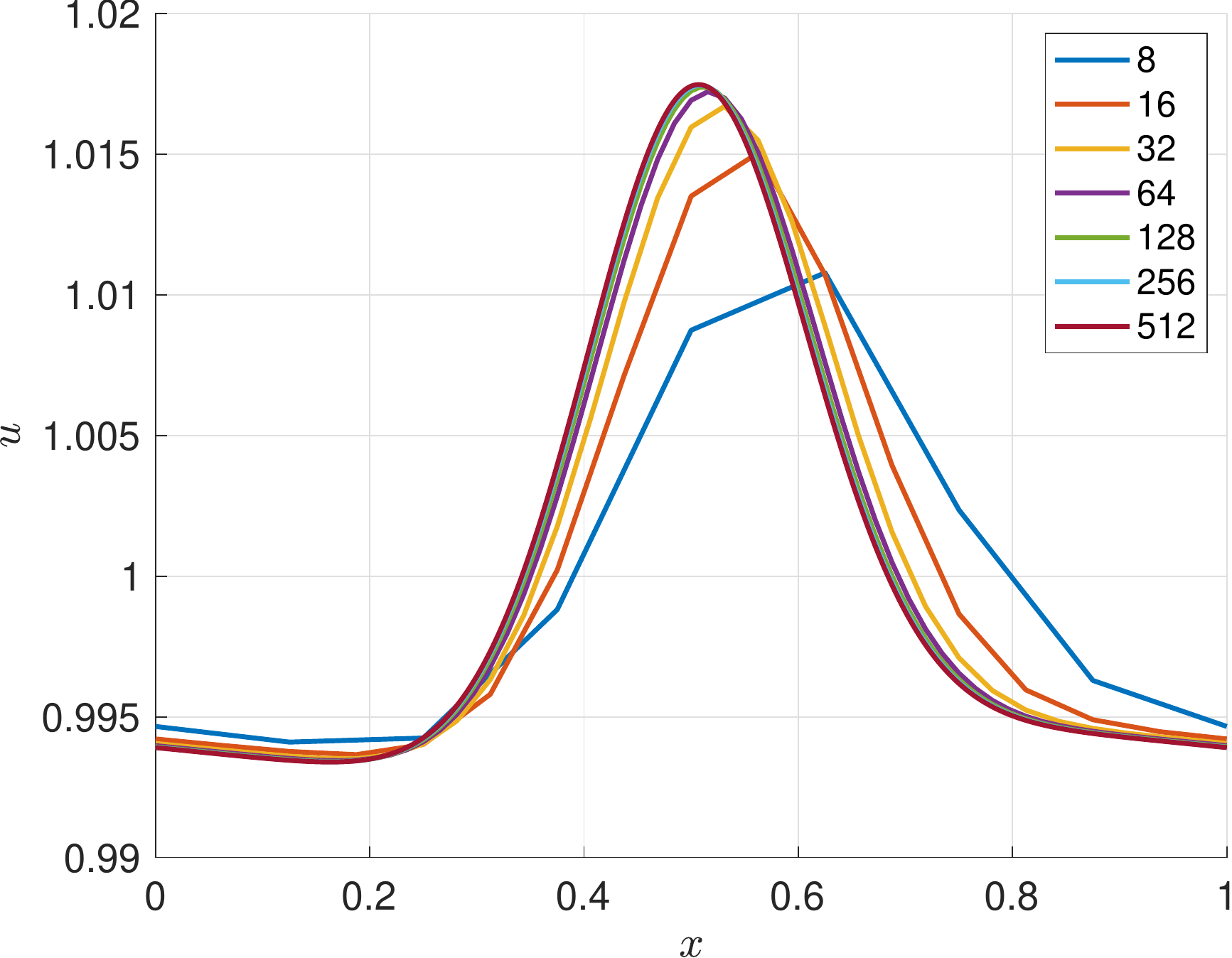} 
        \caption{~}
        \label{fig:steady_ridge_vel_sagy}
    \end{subfigure}\hfill
    \begin{subfigure}{0.45\textwidth}
        \centering
        \includegraphics[width=\textwidth]{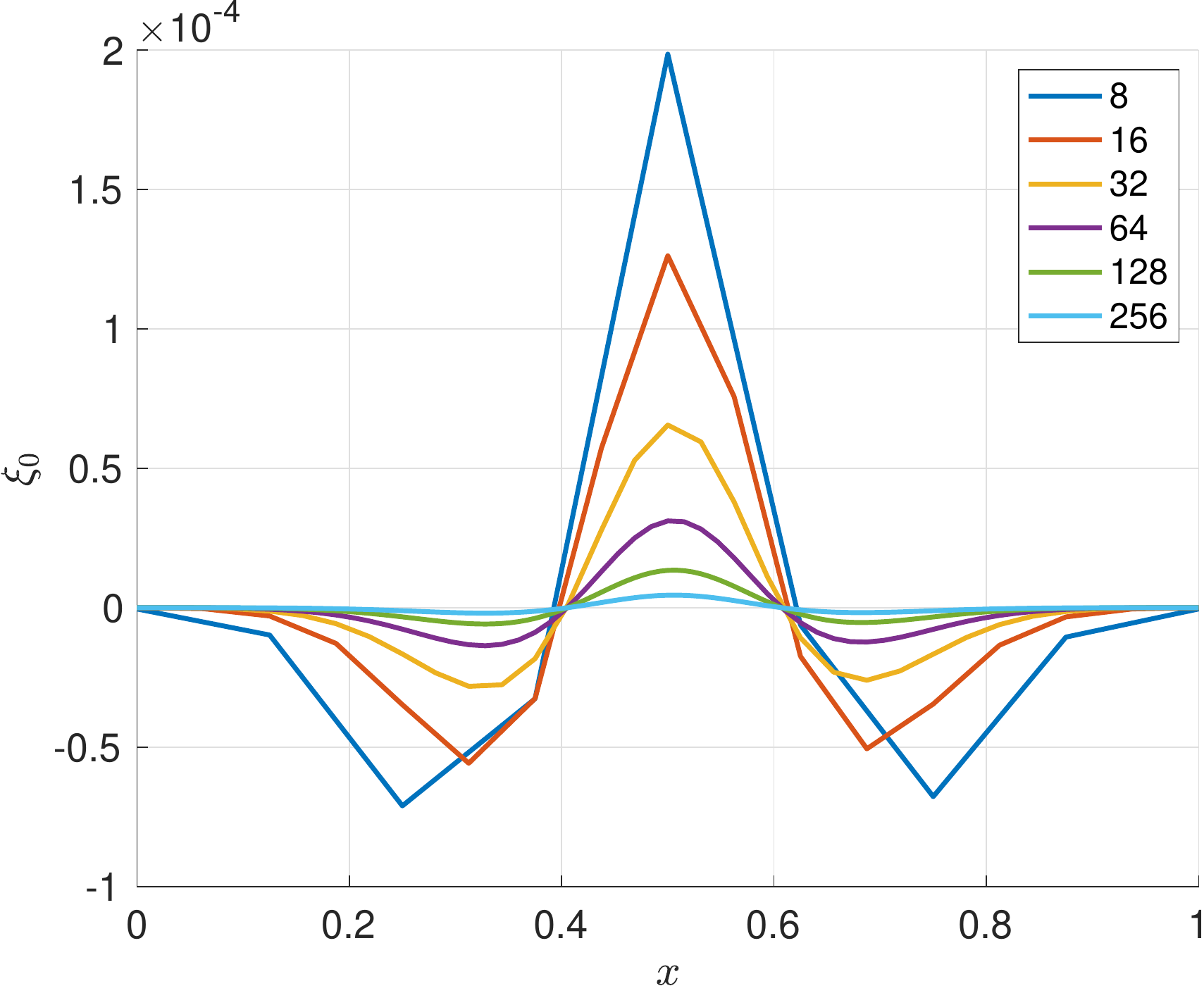} 
        \caption{~}
        \label{fig:steady_ridge_xi0_sagy}
    \end{subfigure}\hfill
    \caption{Left: Steady state of the velocity $u$ for various spatial resolutions using the FD discretization. Right: Time-independent profile $\boldsymbol\xi_0$ as obtained from the EOF algorithm for various spatial resolutions using the FD discretization.}
\end{figure}

\begin{figure}[h!]
    \centering
    \begin{subfigure}{0.45\textwidth}
        \centering
        \includegraphics[width=\textwidth]{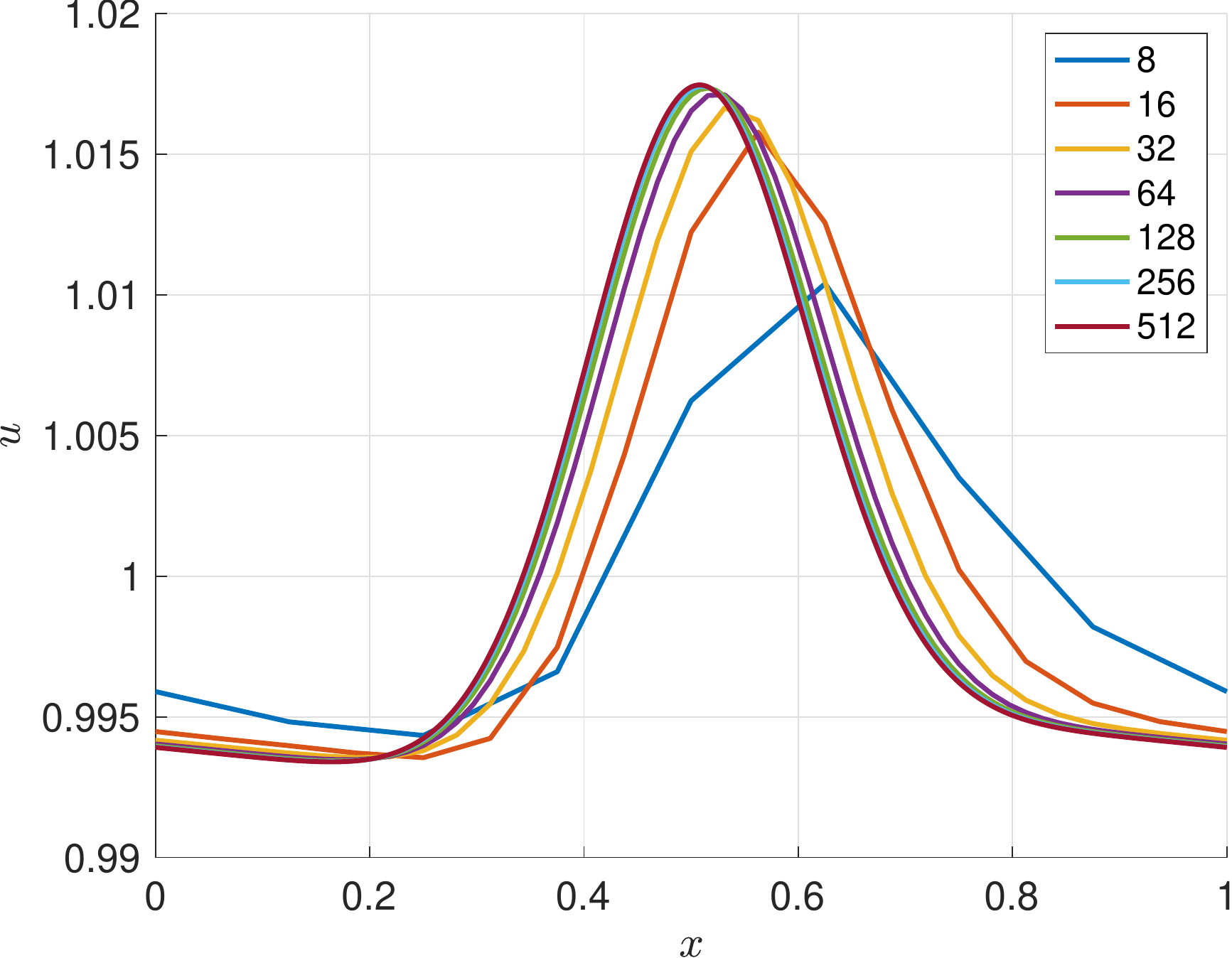} 
        \caption{~}
        \label{fig:steady_ridge_vel_golo}
    \end{subfigure}\hfill
    \begin{subfigure}{0.45\textwidth}
        \centering
        \includegraphics[width=\textwidth]{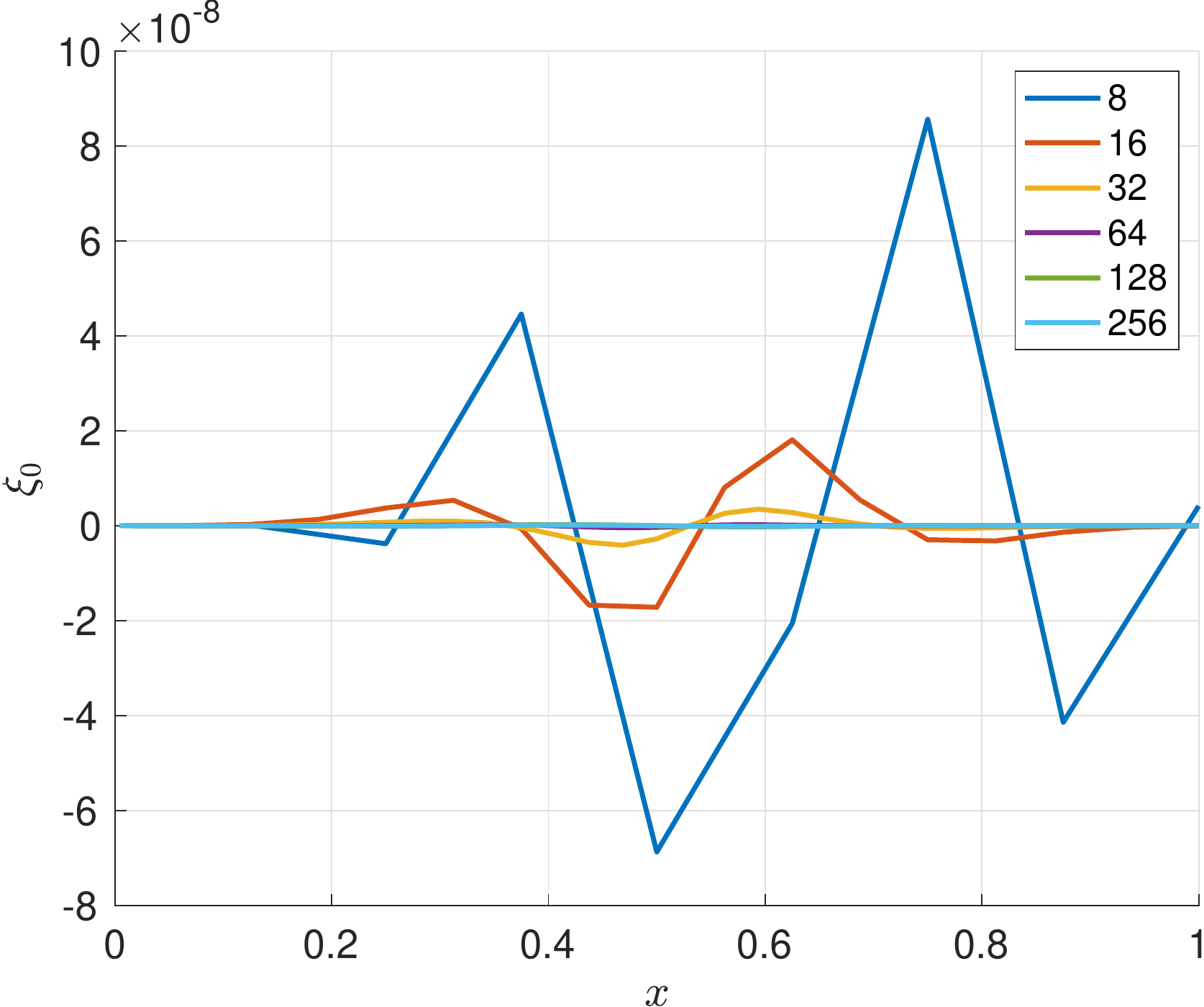} 
        \caption{~}
        \label{fig:steady_ridge_xi0_golo}
    \end{subfigure}\hfill
    \caption{Left: Steady state of the velocity $u$ for various spatial resolutions, using the FE discretization. Right: Time-independent profile $\boldsymbol \xi_0$ as obtained from the EOF algorithm for various spatial resolutions using the FE discretization.}
\end{figure}

The velocities at $t=30$ for various grid sizes are shown in figure \ref{fig:steady_ridge_vel_sagy} for the finite difference discretization and in figure \ref{fig:steady_ridge_vel_golo} for the finite element discretization. 
The corresponding profiles of $\boldsymbol \xi_0$ are reported in figures \ref{fig:steady_ridge_xi0_sagy} and \ref{fig:steady_ridge_xi0_golo}.
For the FD method the dominant error is due to artificial dissipation, associated with the first-order upwind scheme. This error is expected to increase for grid coarsening, as is clearly visible in figure \ref{fig:steady_ridge_xi0_sagy}. Additionally, $\boldsymbol\xi_0$ does not undergo a qualitative change as the grid is coarsened, only increasing in magnitude is observed to attain its largest value where the second derivative of the true velocity is at its highest, indicating that $\boldsymbol \xi_0$ captures the effect of energy dissipation.

The measured errors for the FE method are illustrated in figure \ref{fig:steady_ridge_xi0_golo} and show a neat difference compared to the FD results. 
The FE error is several orders of magnitude smaller than the FD error and is growing in the direction of the flow, which suggests a dispersive-type error.

\begin{figure}[h!]
	\centering
	\includegraphics[width=0.45\textwidth]{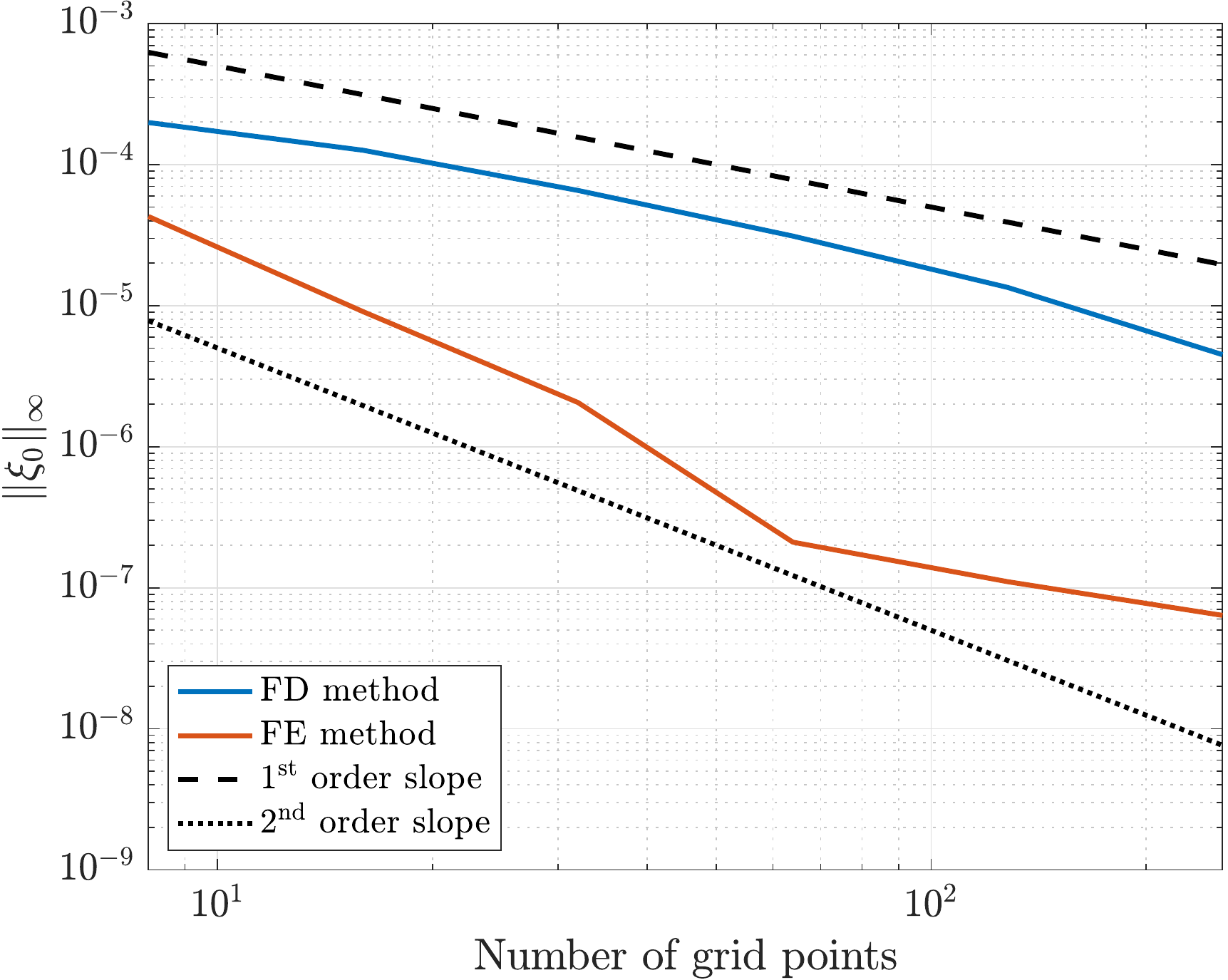}
	\caption{Infinity norm of $\boldsymbol \xi_0$ for various spatial resolutions, for the FD method and the FE method. The dashed and dotted lines depict the slopes for first-order and second-order convergence, respectively.}
	\label{fig: ConvergenceSteadyXi0}
\end{figure}

The orders of convergence of the amplitude of $\boldsymbol \xi_0$ are found to reflect the expected order of accuracy of the methods. This is shown in figure \ref{fig: ConvergenceSteadyXi0}. Using the FD method, $\boldsymbol \xi_0$ shows first-order convergence, second-order convergence is observed for the FE method.


\subsection{Periodic sloshing over a periodic ridge}
\label{subsec:testcase2}
In the second test case time-periodic forcing is applied. This ensures that the velocity does not reach a steady state making this case suitable for analysing the eigenfunctions $\boldsymbol \xi_i$ and their corresponding temporal coefficients $\alpha_i(t)$.
The forcing consists is defined as follows: 
\begin{equation}
    a(t)=C\sum_{j=1}^l n_j\cos\left(\frac{2\pi t}{n_j}\right).
\end{equation}
Here $n_j$ denotes the $j^\mathrm{th}$ mode, with corresponding period $n_j$ and $l$ denotes the number of used modes. The parameter $C$ can be chosen freely and affects the forcing amplitude. We have chosen a value of $C=1/15$ along with the low-frequency forcing term using $n=10$ and high-frequency forcing terms using $n=2$ and $n=1$, respectively. The low-frequency component affects the solution on a long time scale and is the dominant forcing term. The high-frequency components are small disturbances affecting the solution on shorter time scales. The dominance of low-frequency components is incorporated by relating the amplitude of the forcing with the frequency of the forcing.

A spin-up time and a measuring time of five low-frequency forcing periods are adopted. It has been verified by comparing different measuring spin-up times and interval lengths that the spin-up time and measurement interval are sufficiently long to ensure a reliable measurement acquisition. The data that are obtained from these measurements can be regarded as a training data set.

\begin{figure}[h!]
    \centering
    \begin{subfigure}{0.45\textwidth}
        \centering
        \includegraphics[width=\textwidth]{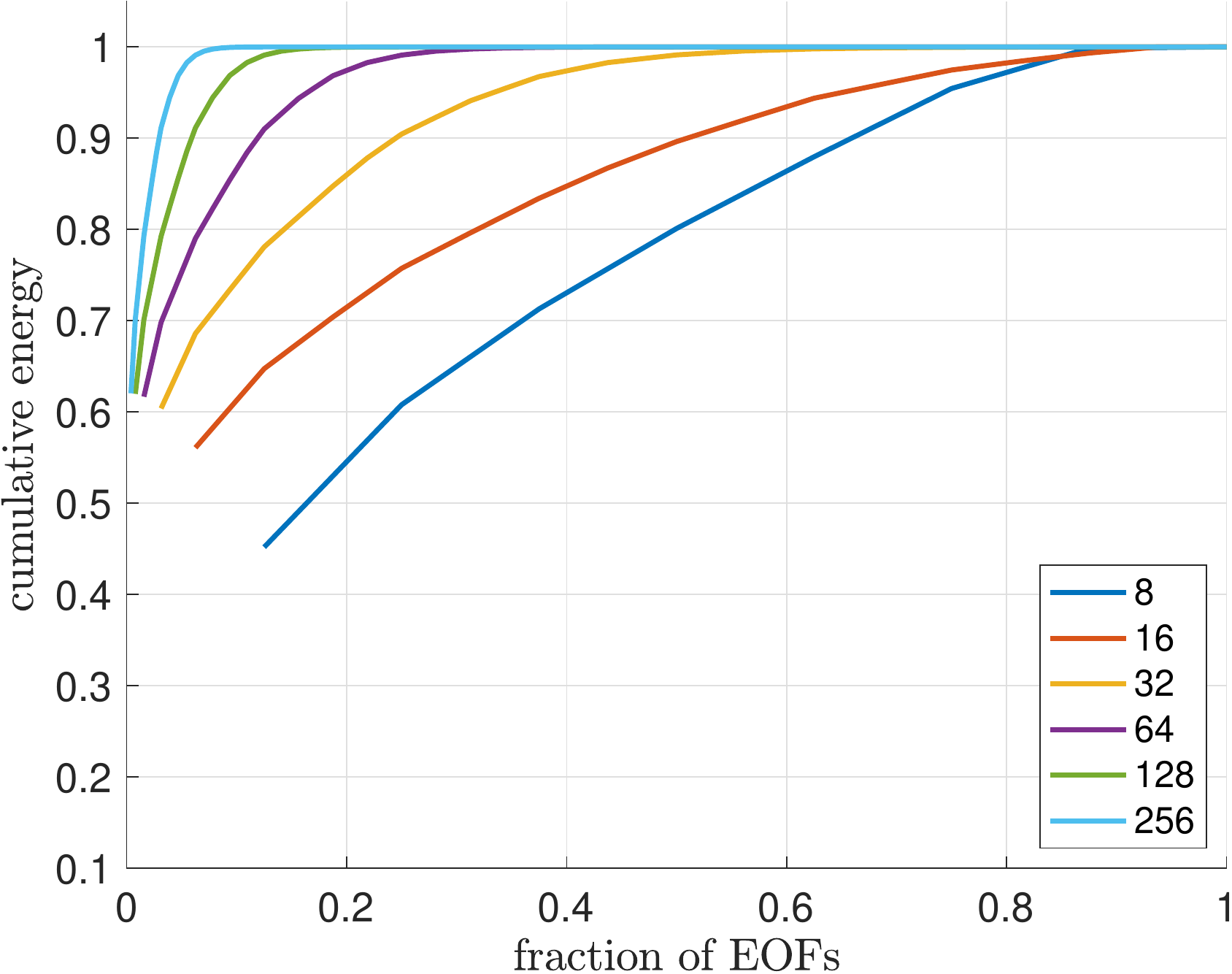} 
        \caption{~}
        \label{fig:EOFspectrum_sagy}
    \end{subfigure} \hfill
    \begin{subfigure}{0.45\textwidth}
        \centering
        \includegraphics[width=\textwidth]{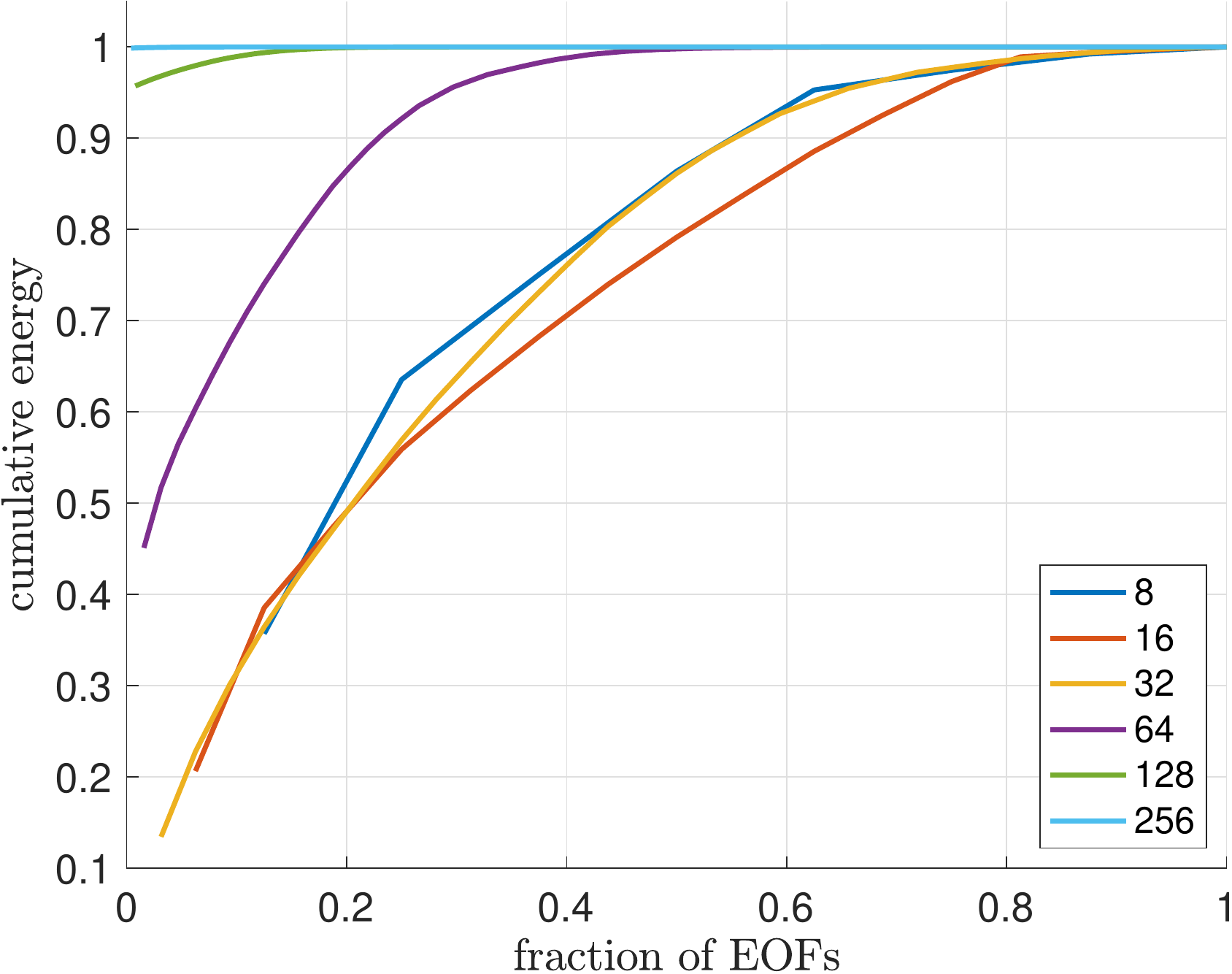} 
        \caption{~}
        \label{fig:EOFspectrum_golo}
    \end{subfigure}
    \caption{Cumulative energy of the subgrid velocity measurements as a function of the number of EOFs for various spatial resolutions, obtained using the FD method (a) and the FE method (b).}
\end{figure}

The eigenvalues corresponding to $\boldsymbol \xi_i$ represent the fraction of energy related to the mode $i$. Of particular interest is the relation between the cumulative energy and the fraction of the available EOFs on various grids. The cumulative energy of $n$ EOFs is given by \begin{equation}
    Q(n)=\frac{\sum_{i=1}^n \lambda_i}{\sum_{k=1}^N \lambda_k},
\end{equation}
where $N$ denotes the total number of EOFs available from the simulation and $\lambda_i$ the eigenvalues.
Figures \ref{fig:EOFspectrum_sagy} and \ref{fig:EOFspectrum_golo} Q as a function of the available EOFs for the FD method and the FE method, respectively. The difference between the truth and the coarse grid simulations decreases as the coarse grids are refined. Correspondingly, the correction toward the truth simulation can be reduced and less of the available data is required to capture the solution's variability. 

Apart from the coarsest grid, the FD method requires the same number of EOF modes to capture nearly all energy of the correction, i.e., with 16 EOFs 99 percent of the variability is captured on all grids with 32 or more grid cells. 
The cumulative energy for the FE method show a markedly different convergence. Almost all variability of the correction on the finest grid is contained within the first EOF, indicating that the coarse-grid solution follows the truth very closely on each of the coarse grids selected. The coarsest solutions each require the same fraction of available EOFs to fully represent the reference solution on the respective grids. A similar result was observed for the cumulative energy of the free surface height. 

\begin{figure}[h!]
    \centering
    \begin{subfigure}{0.45\textwidth}
        \centering
        \includegraphics[width=\textwidth]{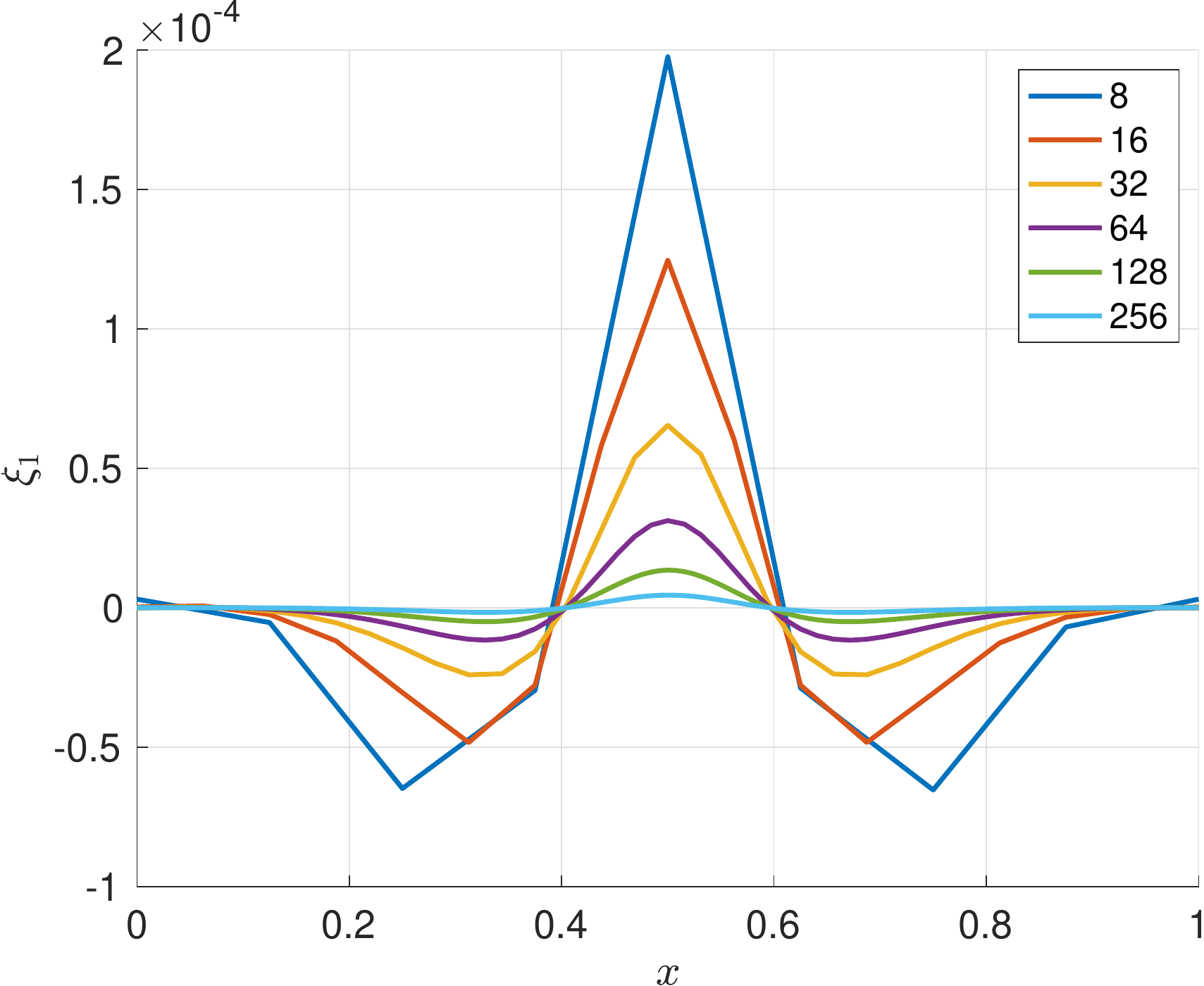}
        \caption{~}
        \label{fig:xi1_sagy}
    \end{subfigure}\hfill
    \begin{subfigure}{0.45\textwidth}
        \centering
        \includegraphics[width=\textwidth]{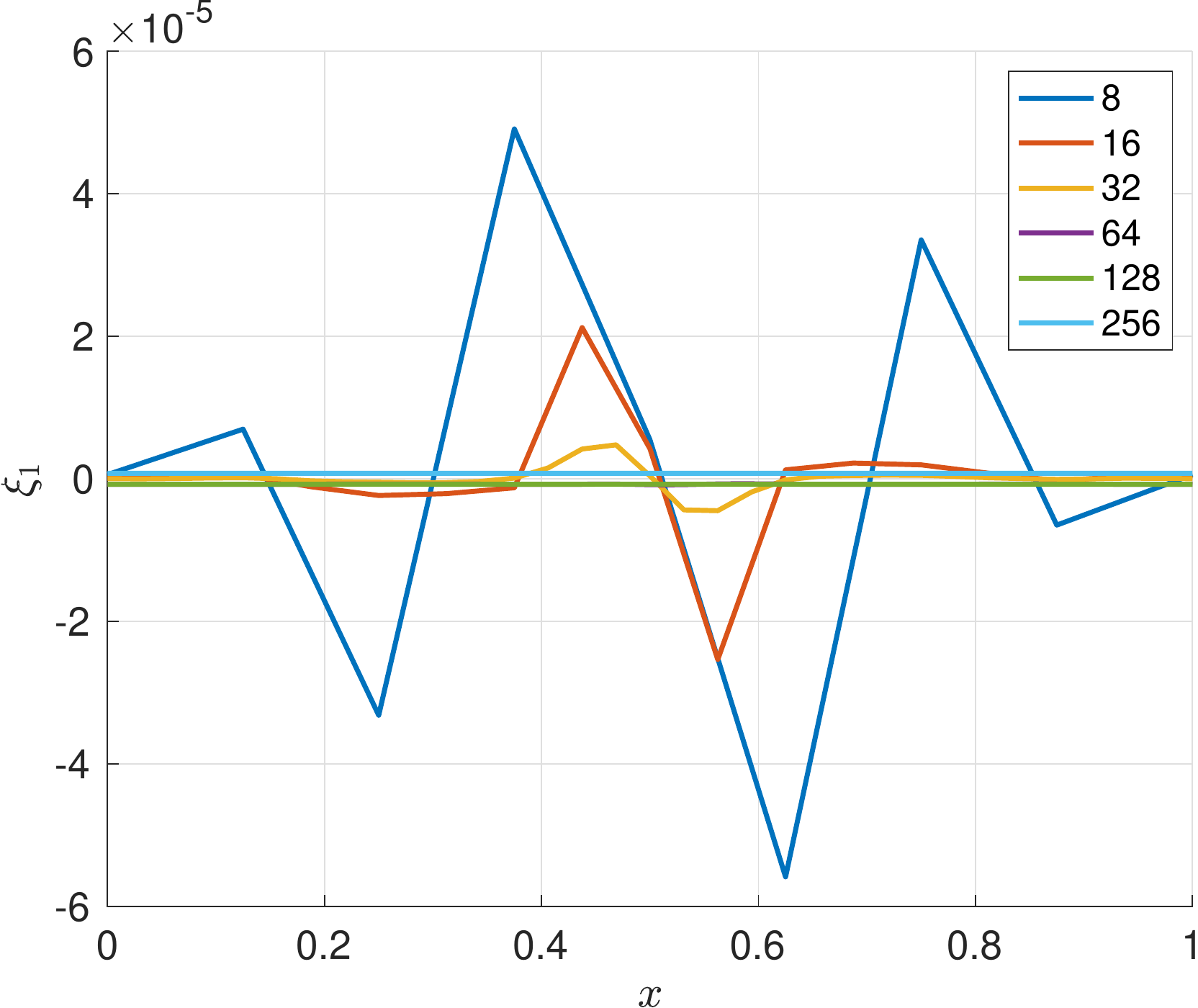} 
        \caption{~}
        \label{fig:xi1_golo}
    \end{subfigure}
    \caption{First EOF mode $\boldsymbol \xi_1$ of the velocity measurements, for various spatial resolutions obtained using the FD method (a) and the FE method (b). The normalized EOF modes have been multiplied by the square root of the corresponding eigenvalues.}
    \label{fig:xi1}
\end{figure}

\begin{figure}[h!]
	\centering
	\begin{subfigure}{0.45\textwidth}
		\centering
		\includegraphics[width=\textwidth]{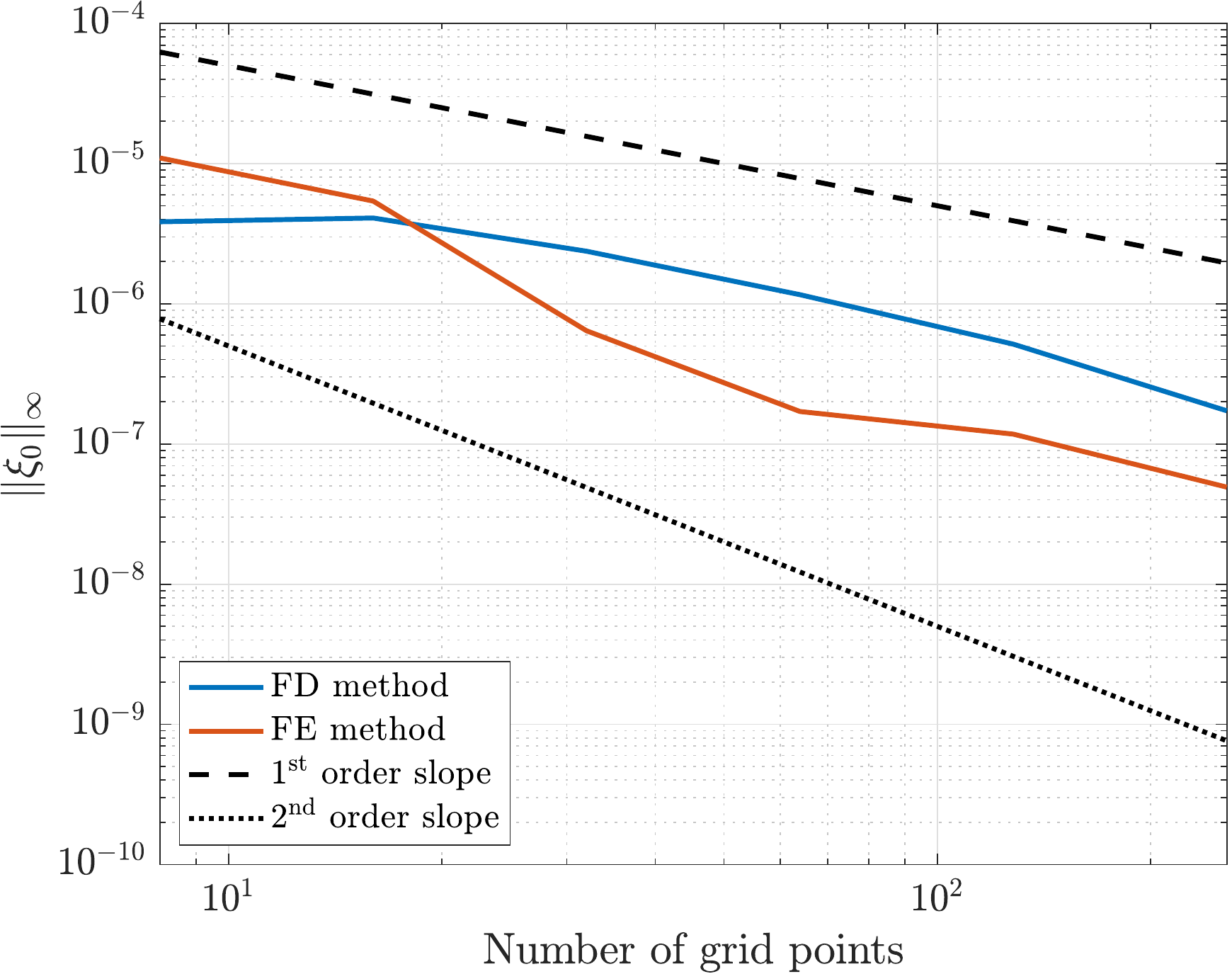}
		\caption{~}
		\label{fig:ConvergenceXi0}
	\end{subfigure} \hfill
	\begin{subfigure}{0.45\textwidth}
		\includegraphics[width=\textwidth]{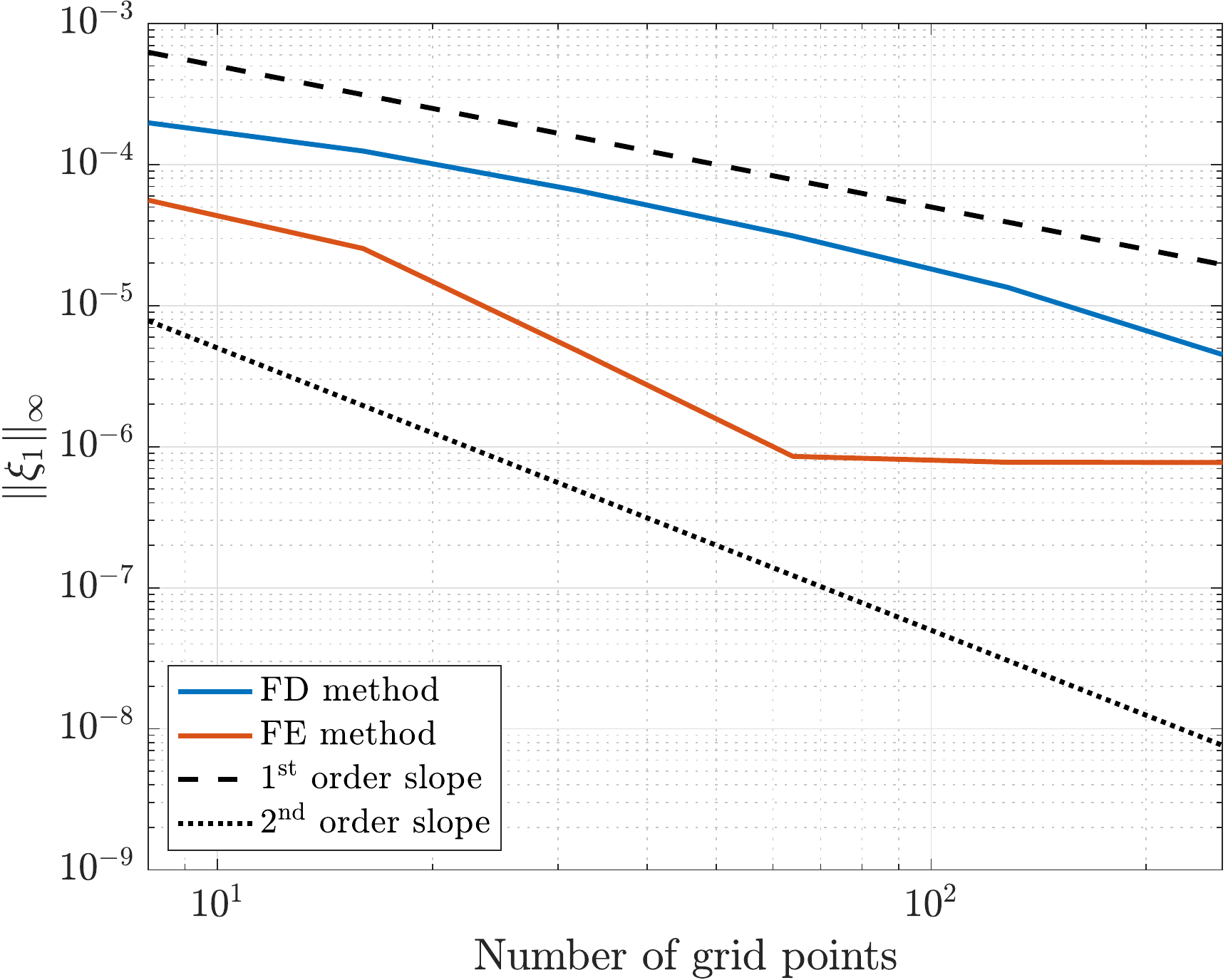}
		\caption{~}
		\label{fig:ConvergenceXi1}
	\end{subfigure}
	\caption{Infinity norm of $\boldsymbol \xi_0$ (a) and $\boldsymbol \xi_1$ (b) for various spatial resolutions, for the FD method and the FE method. The dashed and dotted lines depict the slopes for first-order and second-order convergence respectively.}
	\label{fig:ConvergenceXis}
\end{figure}

In figures \ref{fig:xi1} the first EOF mode for both considered methods is shown. Comparing the different methods, the modes display qualitative differences. The strong difference between the coarsest grid and the finer grids indicates that 8 grid points are too few to resolve the solution of the sloshing problem with the FD method, and hence the captured correction differs strongly from the other computational grids. Convergence of the infinity norm for grid refinement is shown in figure \ref{fig:ConvergenceXis}. The FD method displays first-order convergence and the FE method exhibits second-order convergence. In a similar manner, the EOFs for the free surface height were found to exhibit faster convergence due to the second-order discretization of the continuity equation. First-order convergence was observed for the EOFs for the free surface height.
\begin{figure}[h!]
    \centering
    \begin{subfigure}{0.45\textwidth}
        \centering
        \includegraphics[width=\textwidth]{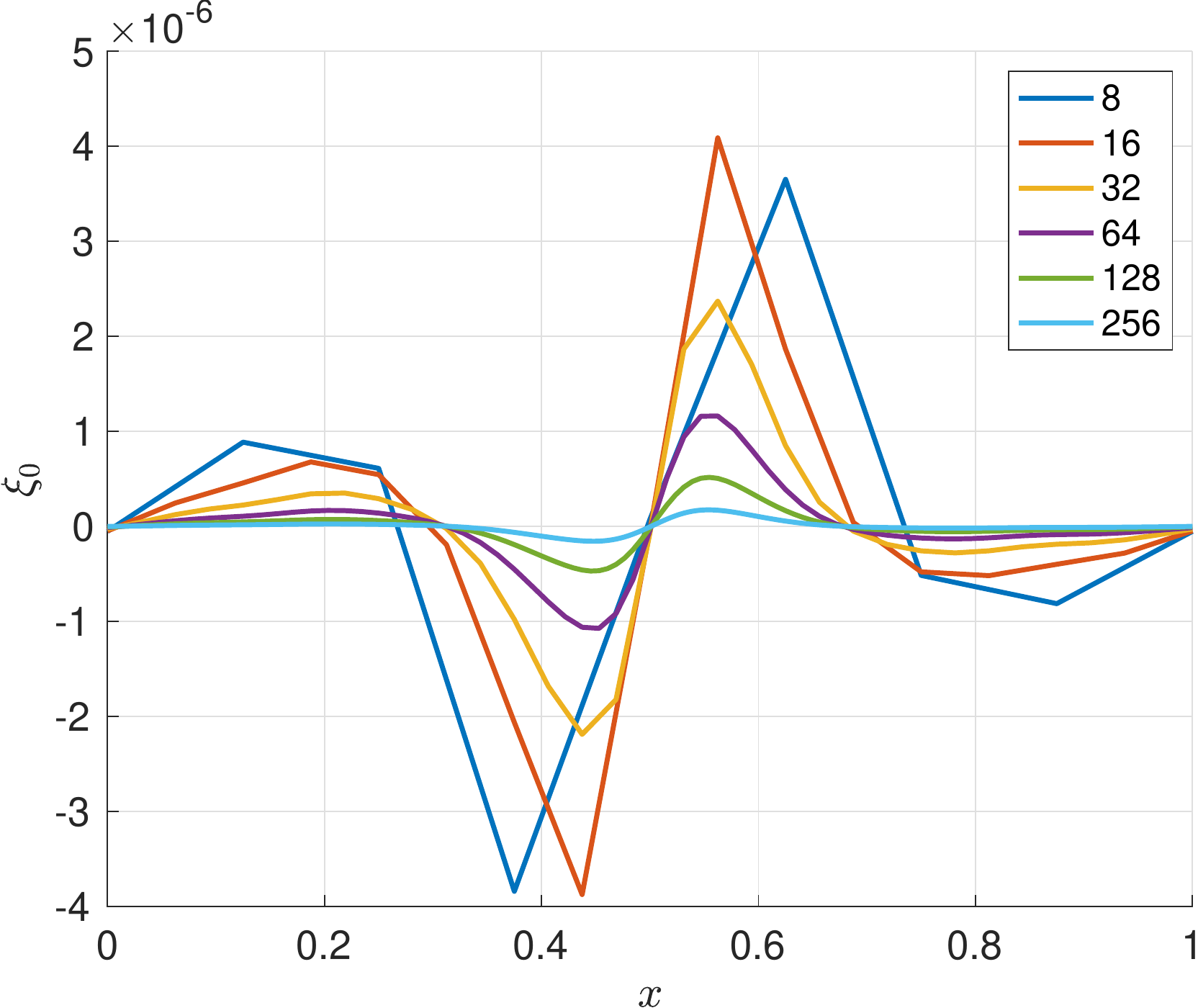} 
        \caption{~}
        \label{fig:sloshing_xi0_sagy}
    \end{subfigure} \hfill
    \begin{subfigure}{0.45\textwidth}
        \centering
        \includegraphics[width=\textwidth]{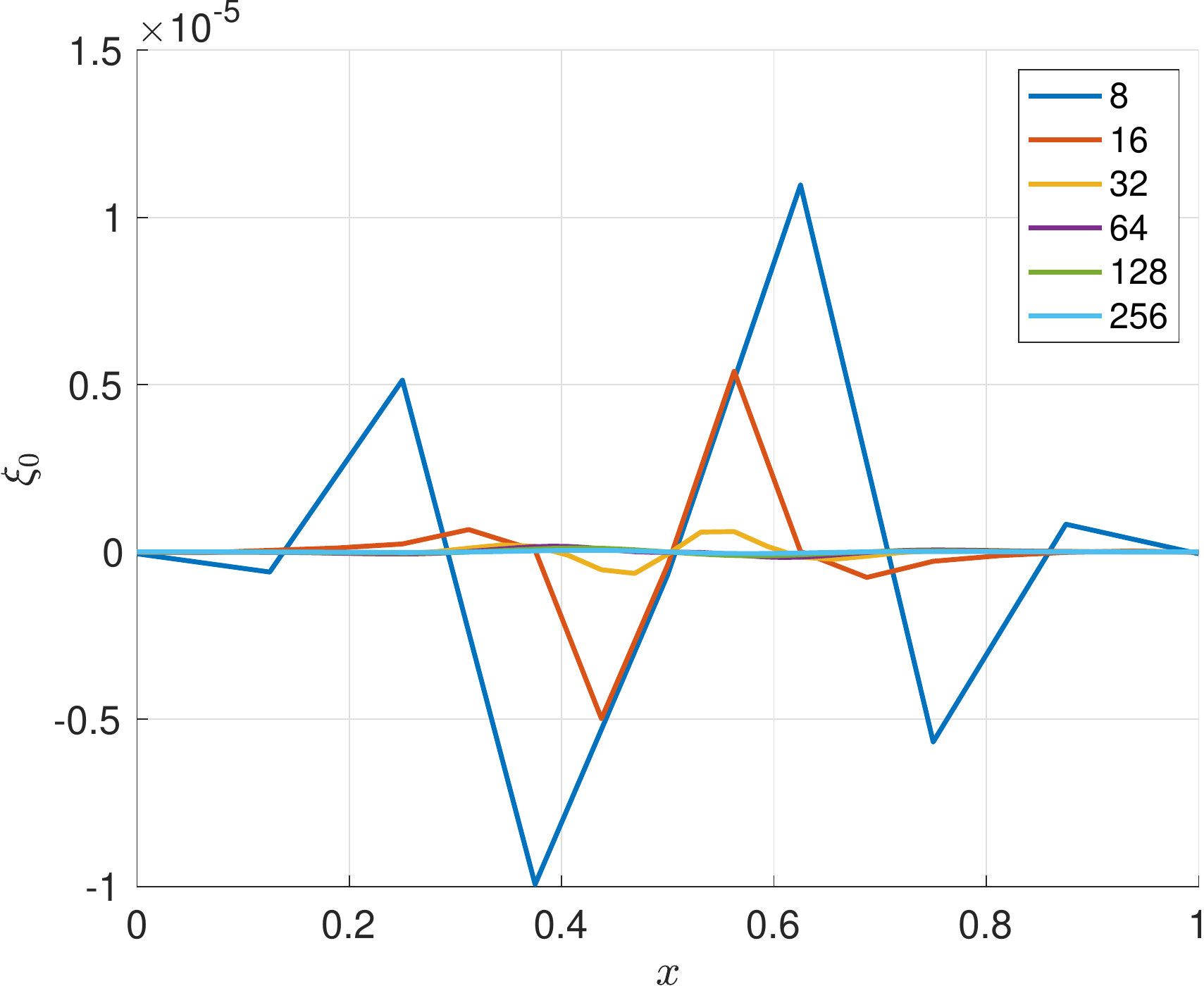} 
        \caption{~}
        \label{fig:sloshing_xi0_golo}
    \end{subfigure}
    \caption{Time-independent profile $\boldsymbol\xi_0$ for the velocity measurements at different grid resolutions using the FD method (a) and the FE method (b).}
\end{figure}

\newpage
\section{Reduced-order corrections based on EOF data}\label{sec:rom}
In this section, we apply the reduced-order model developed in section \ref{subsec:DefiningROM} to the coarse solutions of the previously presented test case with periodic forcing. 

In sections \ref{subsec:ROM_all} and \ref{subsec:ROM_part}, we investigate how the ability of the measured terms to reconstruct the original data set. In these cases, we analyze two grid coarsenings: $32$ and $8$ grid cells. The former resolution allows comparing the FD and FE method for the situation in which they show comparable accuracy, as it was verified numerically. The latter resolution represents a challenging case given the extreme coarsening.  To disentangle the effect of the coarse-grid correction on $u$ and $\eta$, we present the results first for the case in which the reduced-order model is applied to both state variables and then applied to them separately. 

In sections \ref{subsec:ROM_ICs} and \ref{subsec:ROM_statsteadystate} we investigate the robustness of the model for different initial conditions in a periodic regime. Here it is shown that general use of such models requires estimation of the temporal coefficients of the EOFs and how mean quantities might be improved in periodic regimes.


The $L^2$-norm of the pointwise velocity difference with the reference solution is adopted as the error measure, where the reference solution is injected on the coarse grid. Both the FD and the FE discretization use nested grids for the velocity and thus injection is performed trivially. As a measure for the error between the fine and coarse grid solutions we define \begin{equation}
    e(t) = \frac{1}{K+1} \sum_{k=0}^K \left[\sum_{i=1}^N \left(u_\mathrm{truth}(x_i,t+k\Delta t) - u(x_i,t+k\Delta T)\right)^2 \right]^\frac{1}{2},
    \label{eq:ROMerror}
\end{equation}
where $N$ and $x_i$ denote the number and positions of the coarse grid points, respectively.
Time averaging of the error is performed over a period of $K\Delta t$. This time interval is chosen to be one time unit so that the contribution of the high-frequency forcing component to the error remains visible. 


\subsection{Error reduction when correcting all state variables}
\label{subsec:ROM_all}
\begin{figure}[h!]
    \centering
    \begin{subfigure}{0.43\textwidth}
        \centering
        \includegraphics[width=\textwidth]{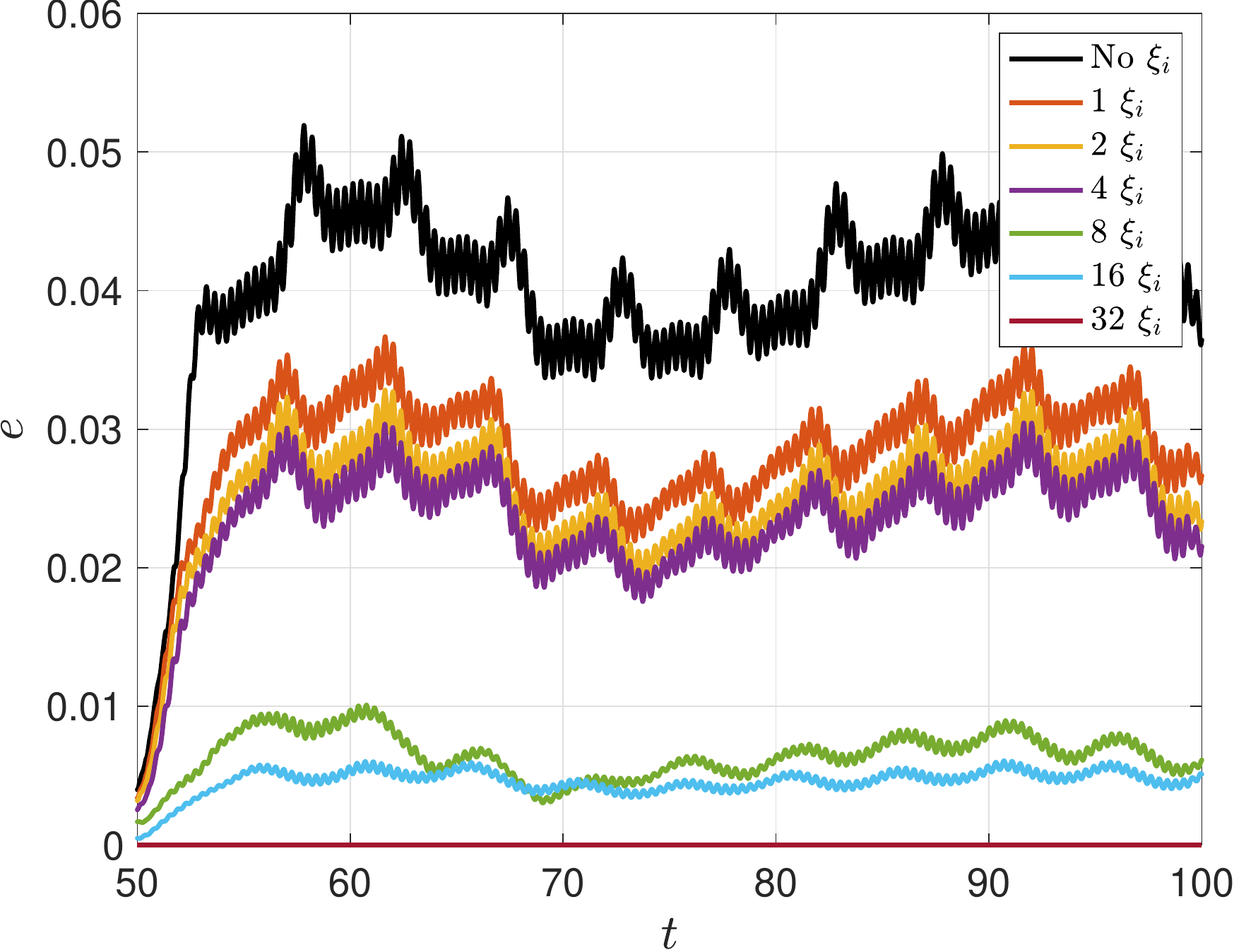}
        \caption{~}
        \label{fig:ROM_sagy}
    \end{subfigure}\hfill
    \begin{subfigure}{0.47\textwidth}
        \centering
        \includegraphics[width=\textwidth]{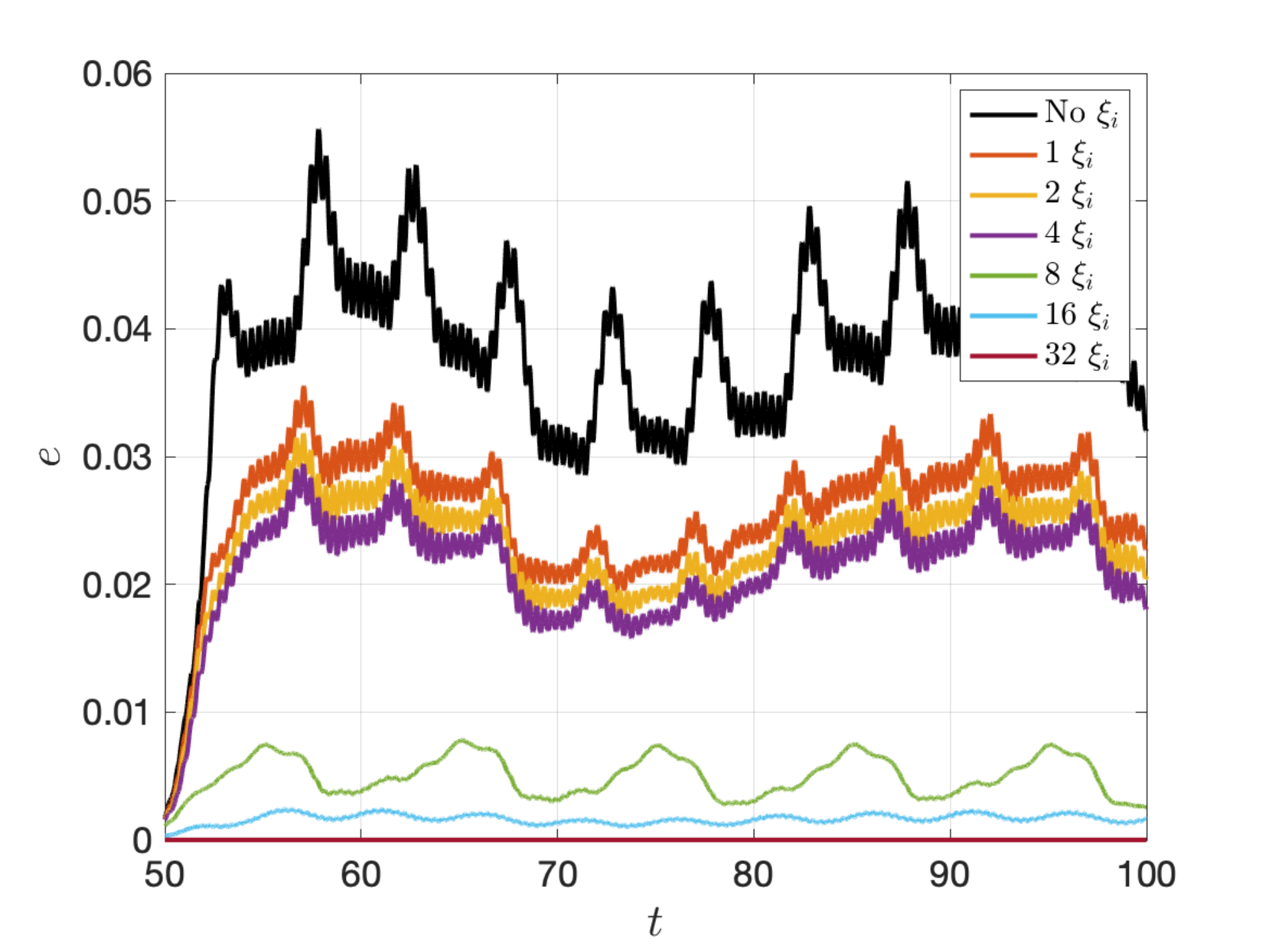} 
        \caption{~}
        \label{fig:ROM_golo}
    \end{subfigure}
    \caption{Error \eqref{eq:ROMerror} on a grid consisting of 32 cells for an increasing number of EOFs using the FD method (a) and the FE method (b). Note that using 32 EOFs recovers the reference solution and zero error is measured.}
\end{figure}

Coarse-grid corrections are applied to both $u$ and $\eta$. Figures \ref{fig:ROM_sagy} and \ref{fig:ROM_golo} show the error reduction over time using an increasing number of EOFs. The mean error values over the time interval $[60,100]$ and the percentage of reduction compared to the coarse solution without correction are given in table \ref{tab:error_reduction32}. Including one EOF in the correction already reduces the error by over 30 percent for both methods. Using a quarter of the available data, 8 out of 32 EOFs, reduces the error by over 80 percent for this test case. 
The computational cost for an increasing number of EOFs for the FD method is given in table \ref{tab: ComputationalCost} and is measured as the CPU time on a local computing cluster. Generating the DNS data takes is the most time consuming part of the algorithm, followed by the computation of the EOFs. No increase in computational cost is observed when including the EOFs in the coarse-grid simulations.

\begin{table}[h!]
    \centering
    \caption{Average values of \eqref{eq:ROMerror} on a grid consisting of 32 cells over the time interval $[60,100]$ as an increasing number of EOFs in included in the coarse-grid correction. The error reduction percentage is calculated with respect to the situation where no correction is applied.}
    \label{tab:error_reduction32}
    \begin{tabular}{|c|c|c|c|c|}
    \hline
    & \multicolumn{2}{c|}{FD} & \multicolumn{2}{c|}{FE}\\
    \hline
     & Mean error & Reduction & Mean error & Reduction \\ 
    \hline
    No correction     & $4.156\times10^{-2}$ & & $3.908\times10^{-2}$ & \\
    1 $\boldsymbol \xi_i$     & $2.898\times10^{-2}$  & 30.2\% & $2.617\times10^{-2}$& 33.0\%\\
    2 $\boldsymbol \xi_i$     & $2.579\times10^{-2}$ & 37.9\% & $2.363\times10^{-2}$& 39.5\%\\
    4 $\boldsymbol \xi_i$     & $2.407\times10^{-2}$ & 42.1\% & $2.162\times10^{-2}$& 44.7\%\\
    8 $\boldsymbol \xi_i$     & $6.343\times10^{-3}$ & 84.7\% & $5.006\times10^{-3}$& 87.2\%\\
    16 $\boldsymbol \xi_i$     & $4.730\times10^{-3}$ & 88.6\% & $1.662\times10^{-13}$& 95.7\%\\
    32 $\boldsymbol \xi_i$     & $2.005\times10^{-13}$ & 100\% & $2.617\times10^{-13}$& 100\%\\
    \hline
    \end{tabular}
\end{table} 
\begin{table}
	\centering
	\caption{Computational cost in seconds for performing the DNS, applying the EOF algorithm and performing coarse-grid simulations with an increasing number of EOFs.}
	\label{tab: ComputationalCost}
	\begin{tabular}{|c|c|}
	\hline
	& Cost \\
	\hline
	DNS & 7.579 \\
	EOF algorithm & 2.652 \\
	Coarse grid, no correction & 0.1599 \\
	Coarse grid, 1 $\boldsymbol \xi_i$ & 0.1522 \\
	Coarse grid, 2 $\boldsymbol \xi_i$ & 0.1535 \\
	Coarse grid, 4 $\boldsymbol \xi_i$ & 0.1521 \\
	Coarse grid, 8 $\boldsymbol \xi_i$ & 0.1547 \\
	Coarse grid, 16 $\boldsymbol \xi_i$ & 0.1519 \\
	Coarse grid, 32 $\boldsymbol \xi_i$ & 0.1519 \\
	\hline
	\end{tabular}
\end{table}

\begin{figure}[h!]
    \centering
    \begin{subfigure}{0.43\textwidth}
        \centering
        \includegraphics[width=\textwidth]{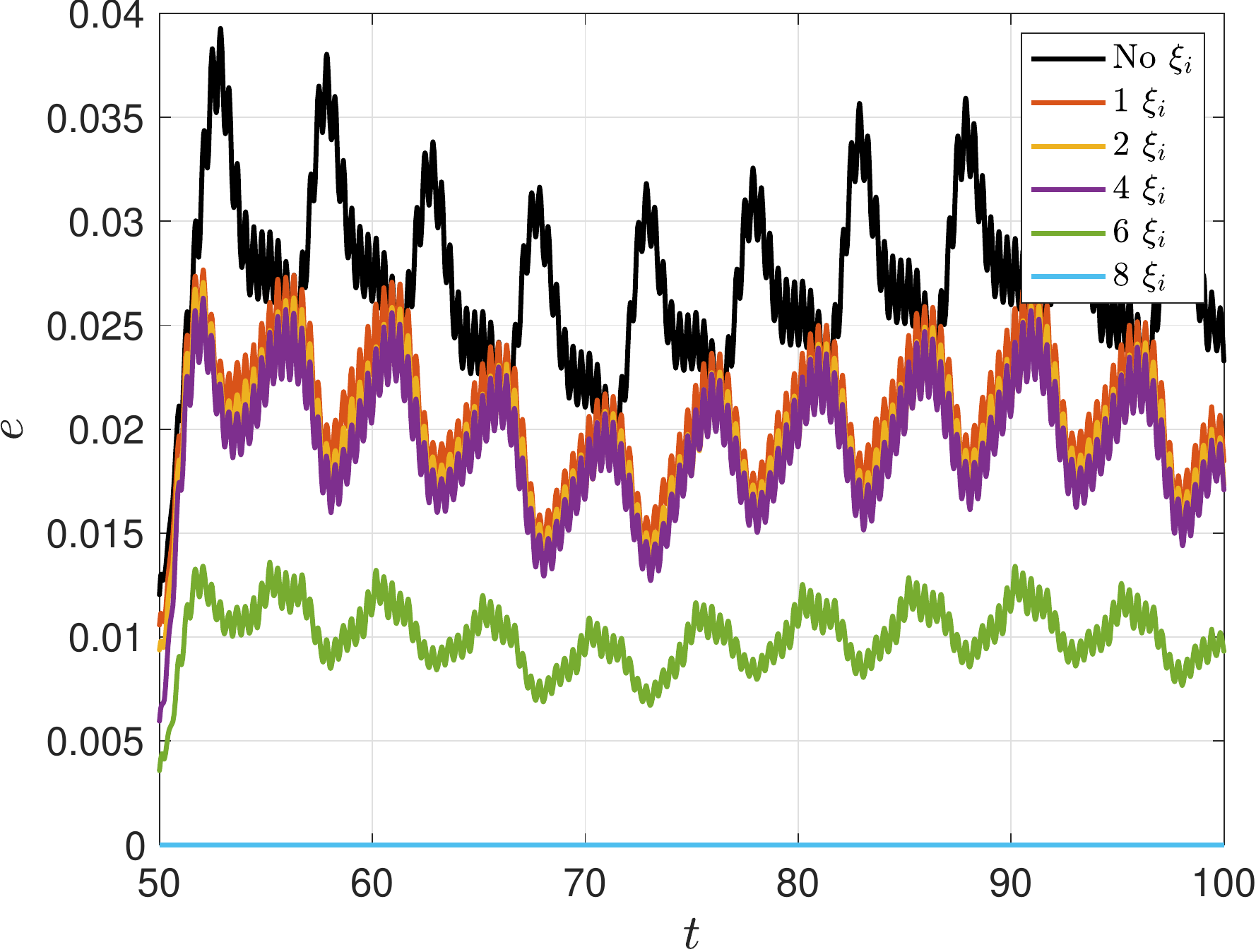}
        \caption{~}
        \label{fig:coarseROM_sagy}
    \end{subfigure}\hfill
    \begin{subfigure}{0.47\textwidth}
        \centering
        \includegraphics[width=\textwidth]{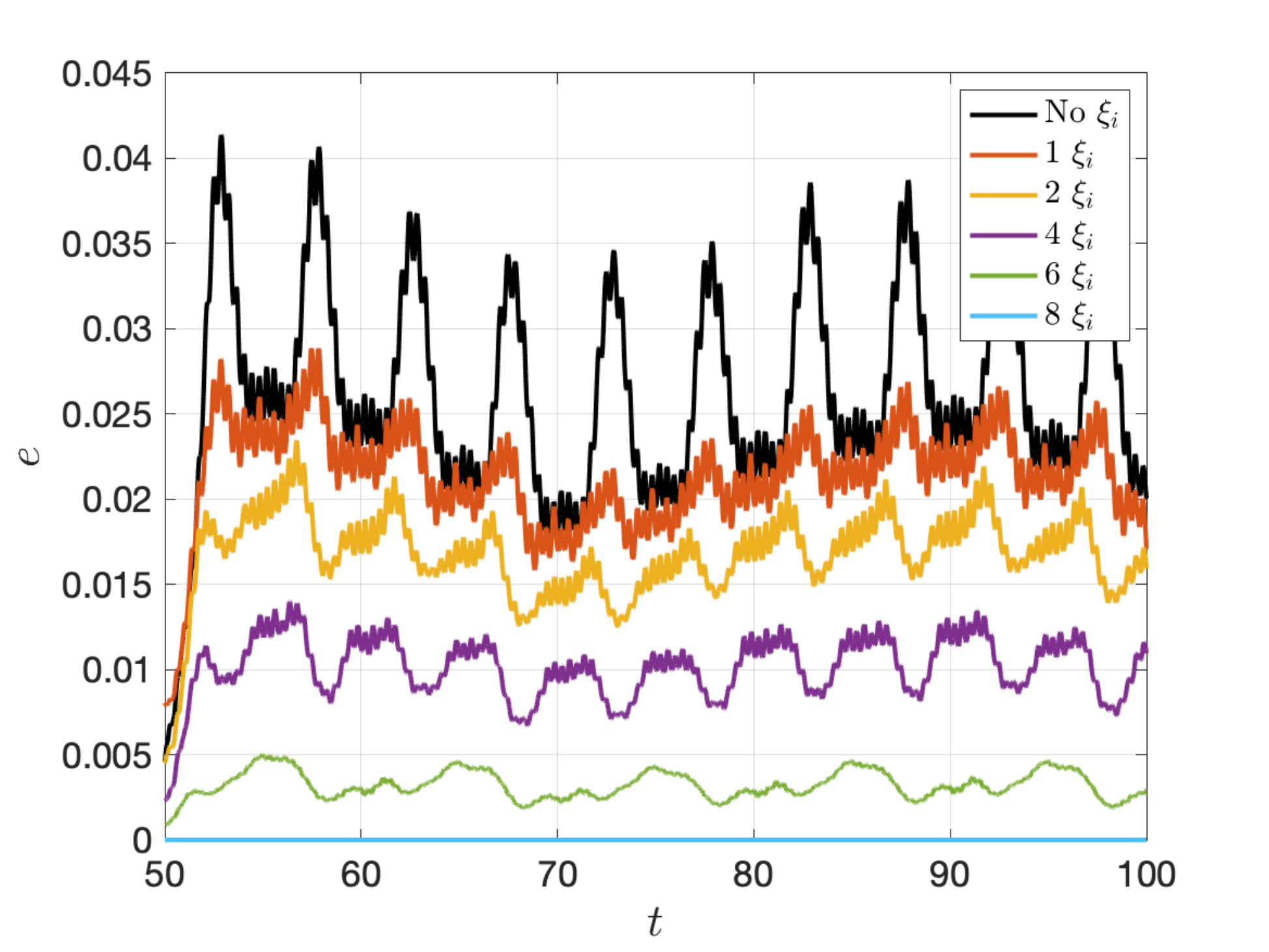} 
        \caption{~}
        \label{fig:coarseROM_golo}
    \end{subfigure}
    \caption{Error \eqref{eq:ROMerror} on a grid consisting of 8 cells for an increasing number of EOFs using the FD method (a) and the FE method (b). Note that using 8 EOFs recovers the reference solution and zero error is measured.}
\end{figure}

Figures \ref{fig:coarseROM_sagy} and \ref{fig:coarseROM_golo} illustrate the error reduction for both methods performed on a grid with 8 computational cells. The method of correction follows from the same principle as shown for 32 cells, but very coarse grids do not allow for an accurate resolution of bathymetry and hence the dynamics of the numerical solution can be vastly different than that of the DNS. The best obtainable result is then achieved by accurately representing the largest scales of the solution and doing so with low computational cost is valuable. 

The mean values of the error are provided in table \ref{tab:error_reduction8}. It can be observed from this table that significant error reduction is possible on this grid even when not using all EOFs. For example, using 6 out of 8 available EOFs yields an error reduction of over 60 percent and 80 percent for the FD method and the FE method, respectively.
\begin{table}[h!]
    \centering
    \caption{Average values of \eqref{eq:ROMerror} on a grid consisting of 8 cells over the time interval $[60,100]$ as an increasing number of EOFs in included in the coarse-grid correction. The error reduction percentage is calculated with respect to the situation where no correction is applied.}
    \label{tab:error_reduction8}
    \begin{tabular}{|c|c|c|c|c|}
    \hline
    & \multicolumn{2}{c|}{FD} & \multicolumn{2}{c|}{FE}\\
    \hline
     & Mean error & Reduction & Mean error & Reduction \\ 
    \hline
    No correction     & $2.675\times10^{-2}$ & & $2.611\times10^{-2}$ & \\
    1 $\boldsymbol \xi_i$     & $2.053\times10^{-2}$  & 23.3\% & $2.126\times10^{-2}$& 18.6\%\\
    2 $\boldsymbol \xi_i$     & $1.953\times10^{-2}$ & 27.0\% & $1.669\times10^{-2}$& 36.1\%\\
    4 $\boldsymbol \xi_i$     & $1.901\times10^{-2}$ & 28.9\% & $1.015\times10^{-2}$& 61.1\%\\
    6 $\boldsymbol \xi_i$     & $9.922\times10^{-3}$ & 62.9\% & $3.202\times10^{-3}$& 87.7\%\\
    8 $\boldsymbol \xi_i$     & $1.025\times10^{-13}$ & 100\% & $2.104\times10^{-13}$& 100\%\\
    \hline
    \end{tabular}
\end{table}
\subsection{Error reduction when correcting one of the two the state variables}
\label{subsec:ROM_part}
\begin{figure}[h!]
    \centering
    \begin{subfigure}{0.45\textwidth}
        \centering
        \includegraphics[width=\textwidth]{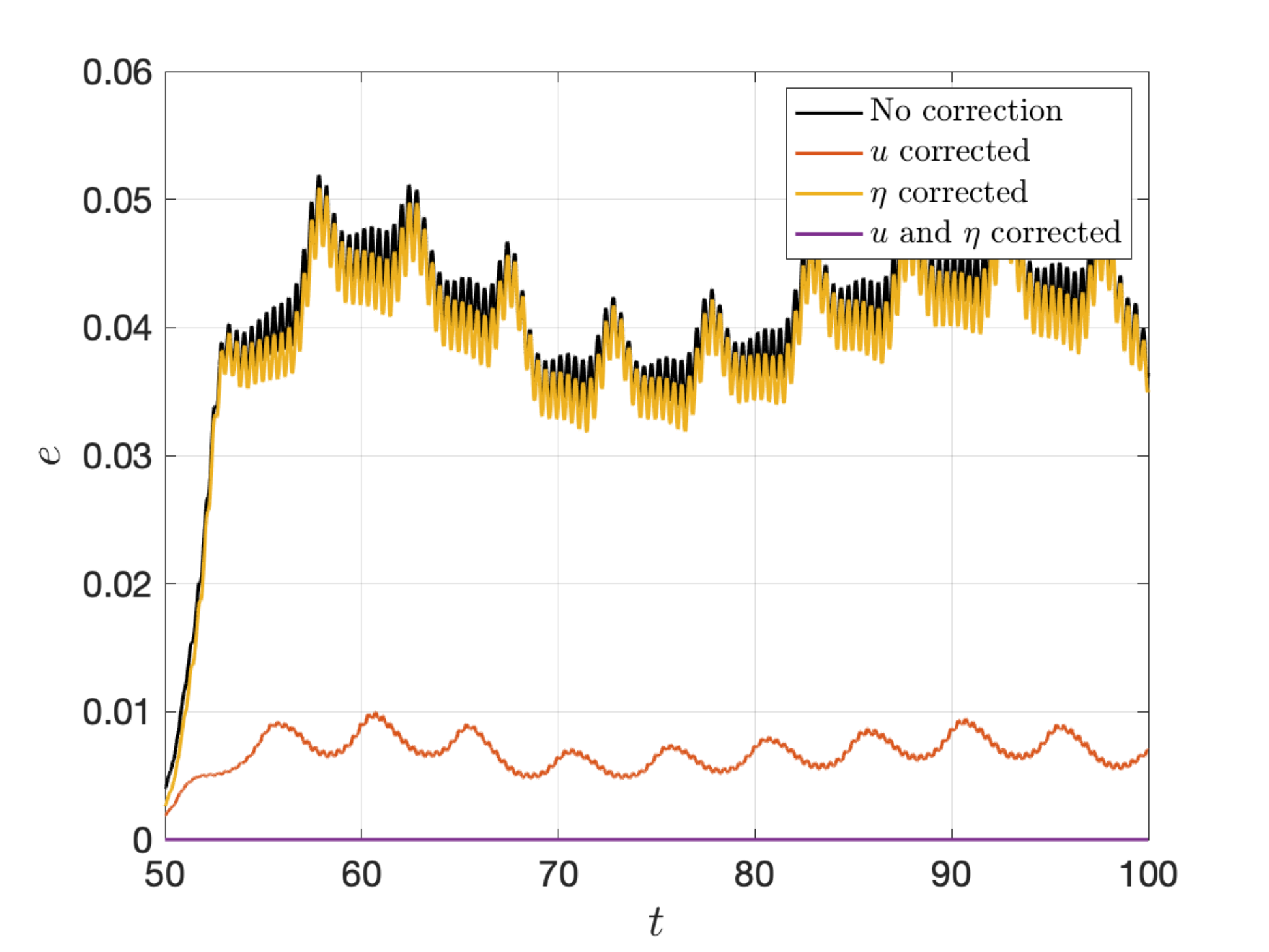}
        \caption{~}
        \label{fig:ROM_improved_sagy}
    \end{subfigure}\hfill
    \begin{subfigure}{0.45\textwidth}
        \centering
        \includegraphics[width=\textwidth]{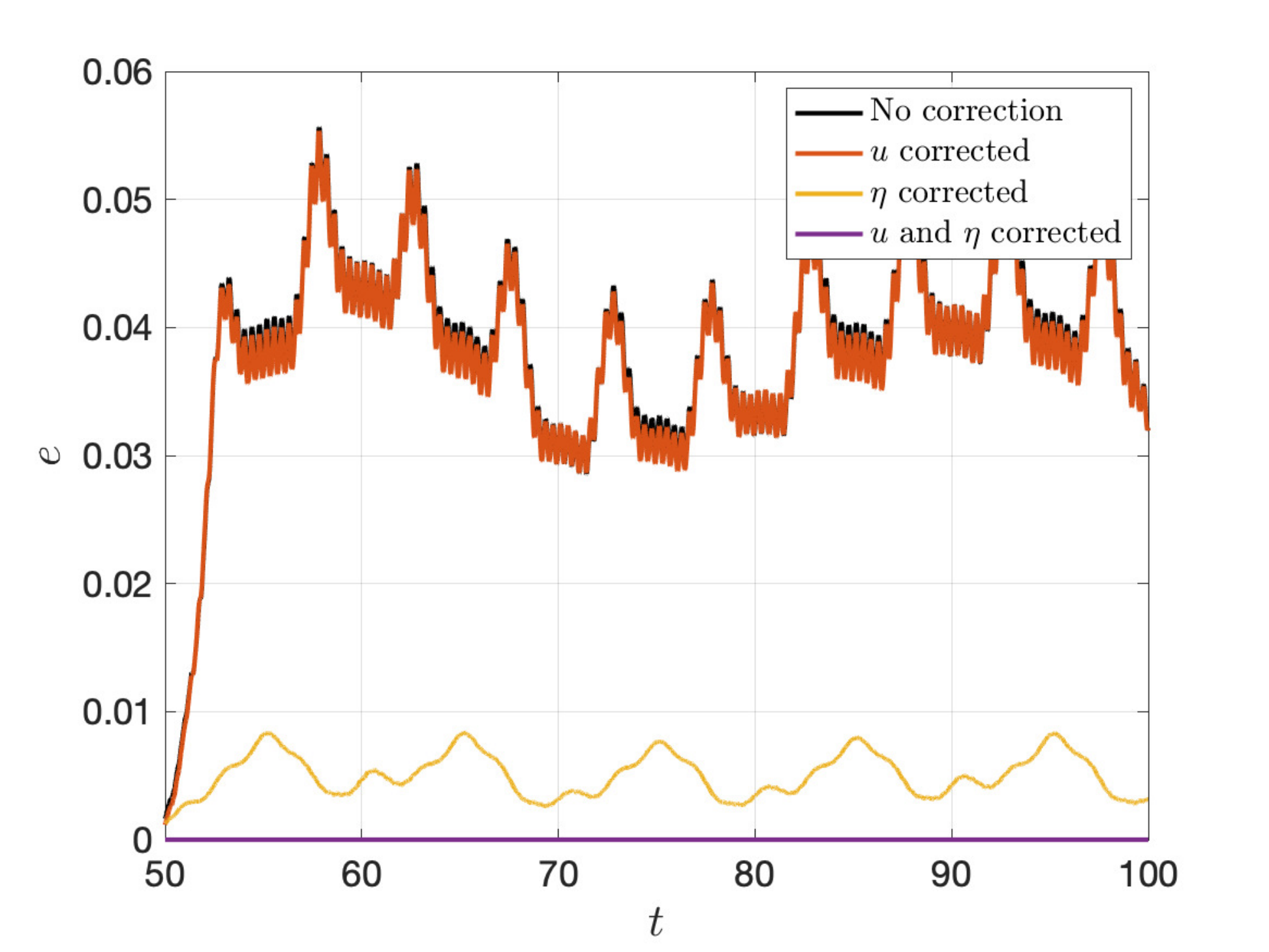} 
        \caption{~}
        \label{fig:ROM_improved_golo}
    \end{subfigure}
    \caption{Error \eqref{eq:ROMerror} on a grid consisting of 32 cells as either the velocity or the free surface height is fully corrected, using the full set of EOFs for the FD method (a) and the FE method (b). Note that correcting both the velocity and the free surface height produces zero error.}
\end{figure}

Figures \ref{fig:ROM_improved_sagy} and \ref{fig:ROM_improved_golo} show the error reduction when only one of the variables is corrected, using the FD method and the FE method, respectively. The coarse grid consists of 32 computational cells for this comparison and coarse-grid corrections are implemented using the full set of computed EOFs for the considered state variable.

The reduced error in figure \ref{fig:ROM_improved_sagy} shows a considerable improvement if the $u$ correction is analyzed. This is in agreement with the fact that the first-order upwind scheme used for convection introduces the dominant source of error. Applying a correction to the free surface height does not yield significant improvement, since the error in the momentum equation dominates. 
Conversely, for the FE method the correction of the momentum equation does not lead to any significant improvement, as the FE method employed here shows high accuracy by itself. As mentioned in section \ref{sec:governingequations}, the FE method adopts first and zeroth order polynomials in the discretization of the momentum equation and continuity equation, respectively. Thus, it is reasonable to expect that correcting the free surface height strongly reduces the overall error since this is the dominant source of error. This is observed in figure \ref{fig:ROM_improved_golo}. 
\begin{table}[h!]
    \centering
    \caption{Average values of \eqref{eq:ROMerror} on a grid consisting of 32 cells over the time interval $[60,100]$ as either the velocity or the free surface height are fully corrected. The error reduction percentage is calculated with respect to the situation where no correction is applied.}
    \label{tab:error_reduction_onevariable}
    \begin{tabular}{|c|c|c|c|c|}
    \hline
    & \multicolumn{2}{c|}{FD} & \multicolumn{2}{c|}{FE}\\
    \hline
     & Mean error & Reduction & Mean error & Reduction \\ 
    \hline
    No correction     & $4.156\times10^{-2}$ & & $3.908\times10^{-2}$ & \\
    $u$ corrected     & $6.888\times10^{-3}$  & 83.4\% & $3.872\times10^{-2}$& 0.870\%\\
    $\eta$ corrected    & $4.014\times10^{-2}$ & 3.27\% & $5.089\times10^{-3}$& 87.0\%\\
    $u$ and $\eta$ corrected    & $2.005 \times10^{-13}$& 100\% & $2.617\times10^{-13}$& 100\%\\
    \hline
    \end{tabular}
\end{table}

\newpage
\subsection{Sensitivity to initial conditions} \label{subsec:ROM_ICs}
In this subsection, we investigate the accuracy of predictions under perturbations of the initial conditions. The aim is to probe the robustness of the model in actual predictions, where the initial condition is in general different from that used in the dataset the model was trained on. By changing the initial conditions, the evolution of the flow is changed and thus the measured time series and EOFs constitute a correction term that no longer coincides with the exact subgrid data. The results are presented for the finite difference method using a reference grid of 512 computational cells and a coarse grid of 32 computational cells.

\begin{figure}[h!]
    \centering
    \begin{subfigure}{0.45\textwidth}
        \centering
        \includegraphics[width=\textwidth]{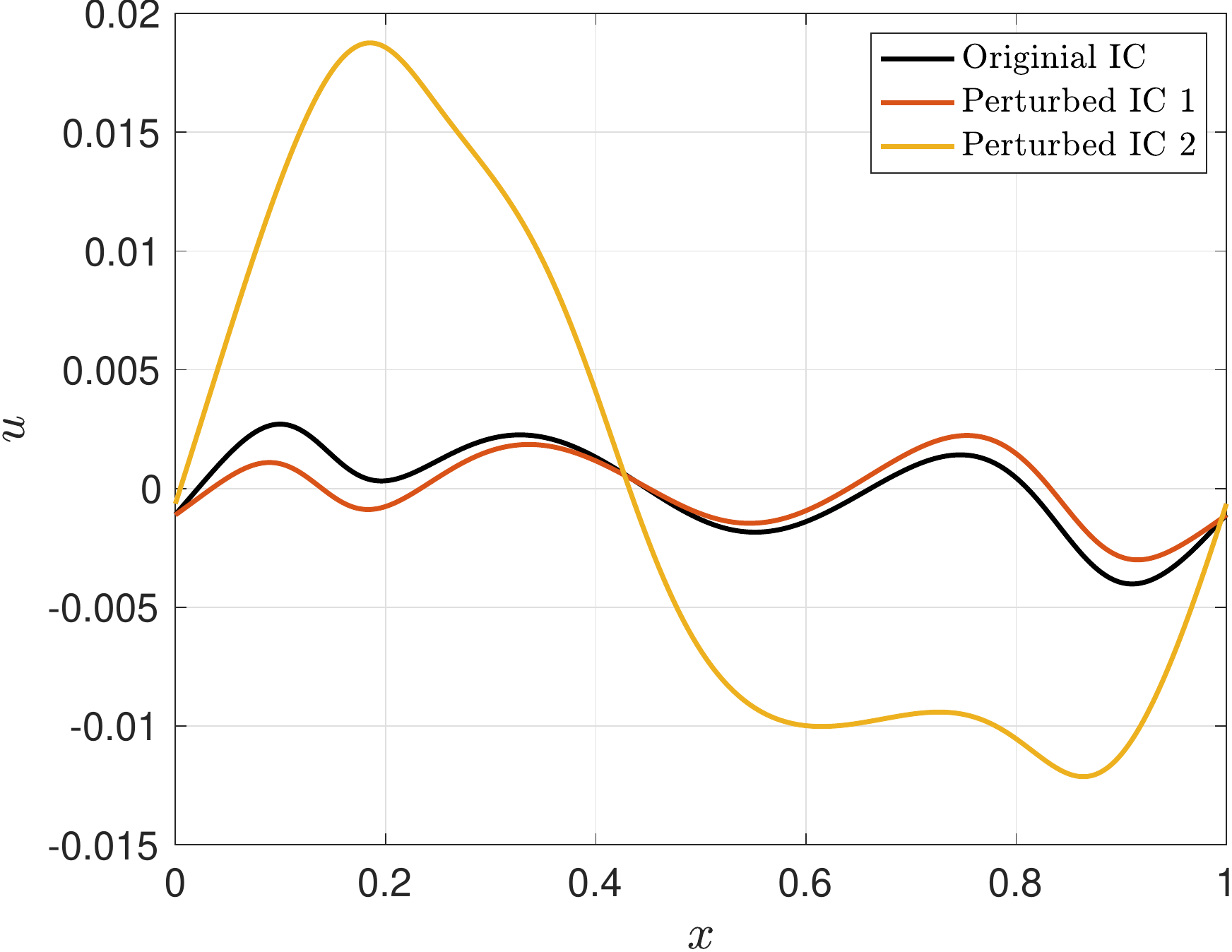}
        \caption{~}
    \end{subfigure}\hfill
    \begin{subfigure}{0.45\textwidth}
        \centering
        \includegraphics[width=\textwidth]{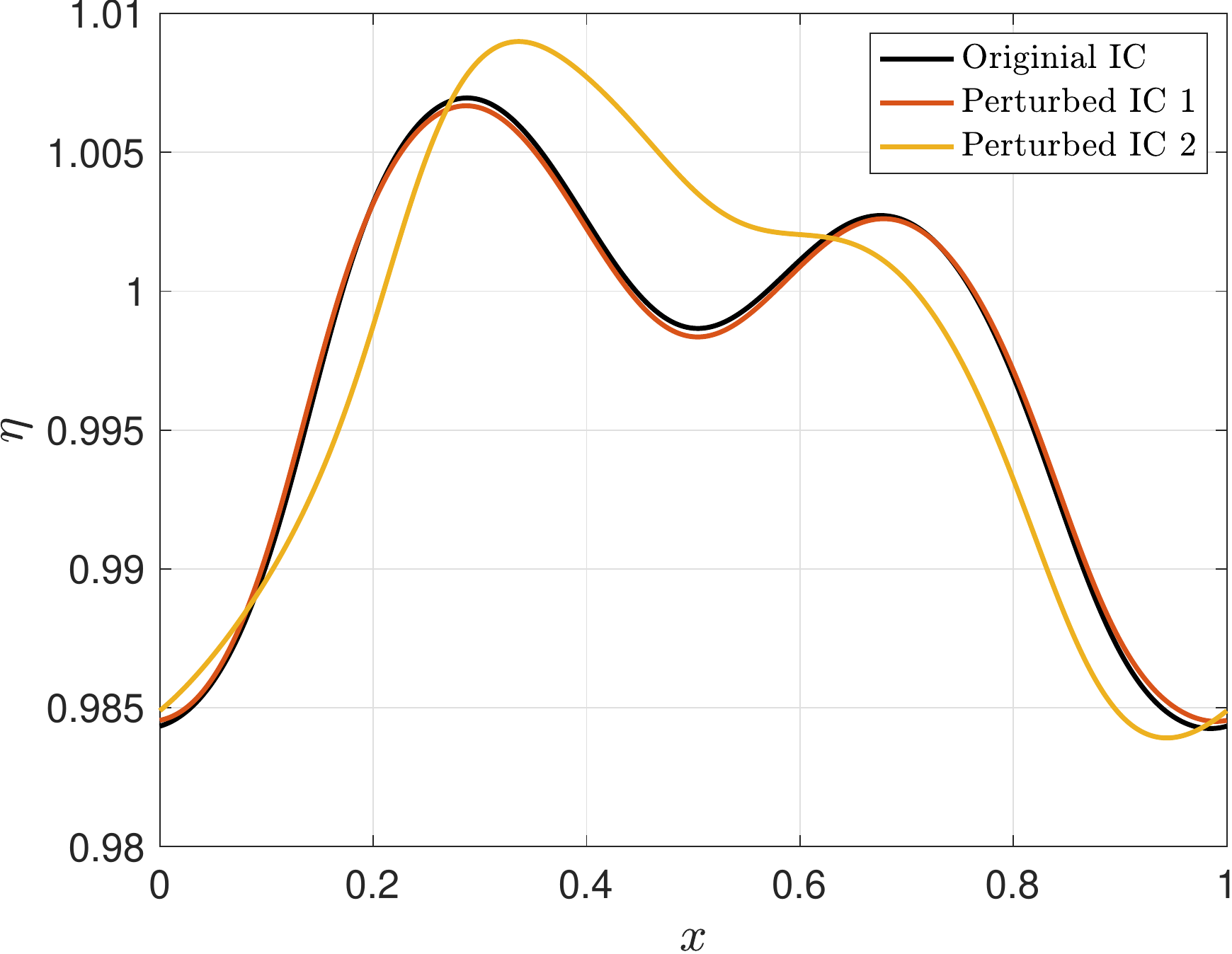} 
        \caption{~}
    \end{subfigure}
    \caption{Initial conditions used to establish the sensitivity of the reduced-order correction term. Both the initial velocities (a) and the initial free surface height profiles (b) are obtained by sampling the numerical solution at specified times. The initial conditions for the original data set is given in black, the red line and yellow lines denote the perturbed initial conditions.}
    \label{fig:perturbedICs}
\end{figure}

The perturbed initial conditions and the initial conditions used to generate the reduced-order corrections are found in figure \ref{fig:perturbedICs}. The perturbed ICs are obtained by sampling the DNS result at times $t=100$ and $t=10$ and are referred to as perturbed IC 1 and perturbed IC 2, respectively. It can be observed that the former slightly deviates from the original initial condition, while the latter deviates significantly.

\begin{figure}[h!]
    \centering
    \begin{subfigure}{0.45\textwidth}
        \centering
        \includegraphics[width=\textwidth]{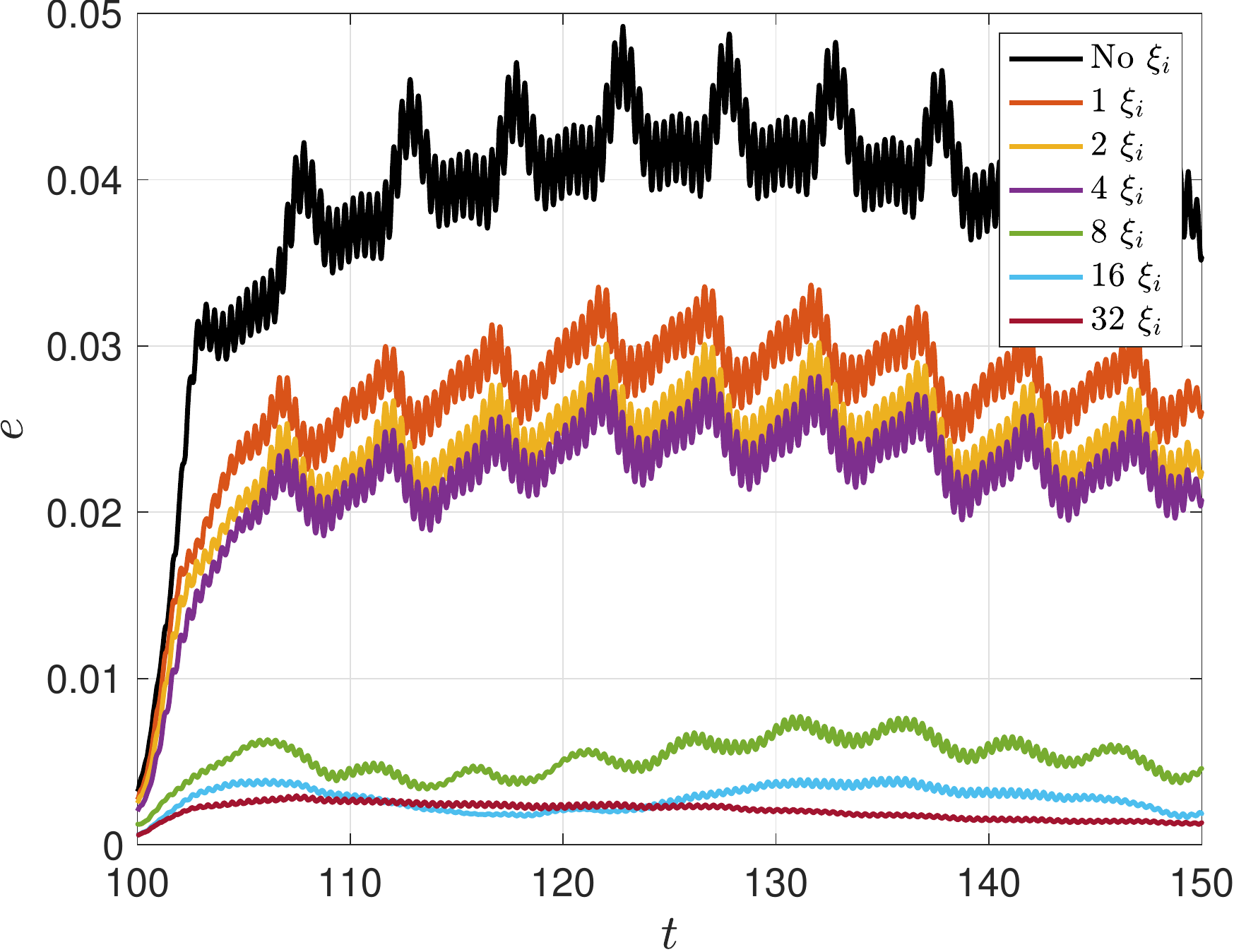}
        \caption{~}
    \end{subfigure}\hfill
    \begin{subfigure}{0.45\textwidth}
        \centering
        \includegraphics[width=\textwidth]{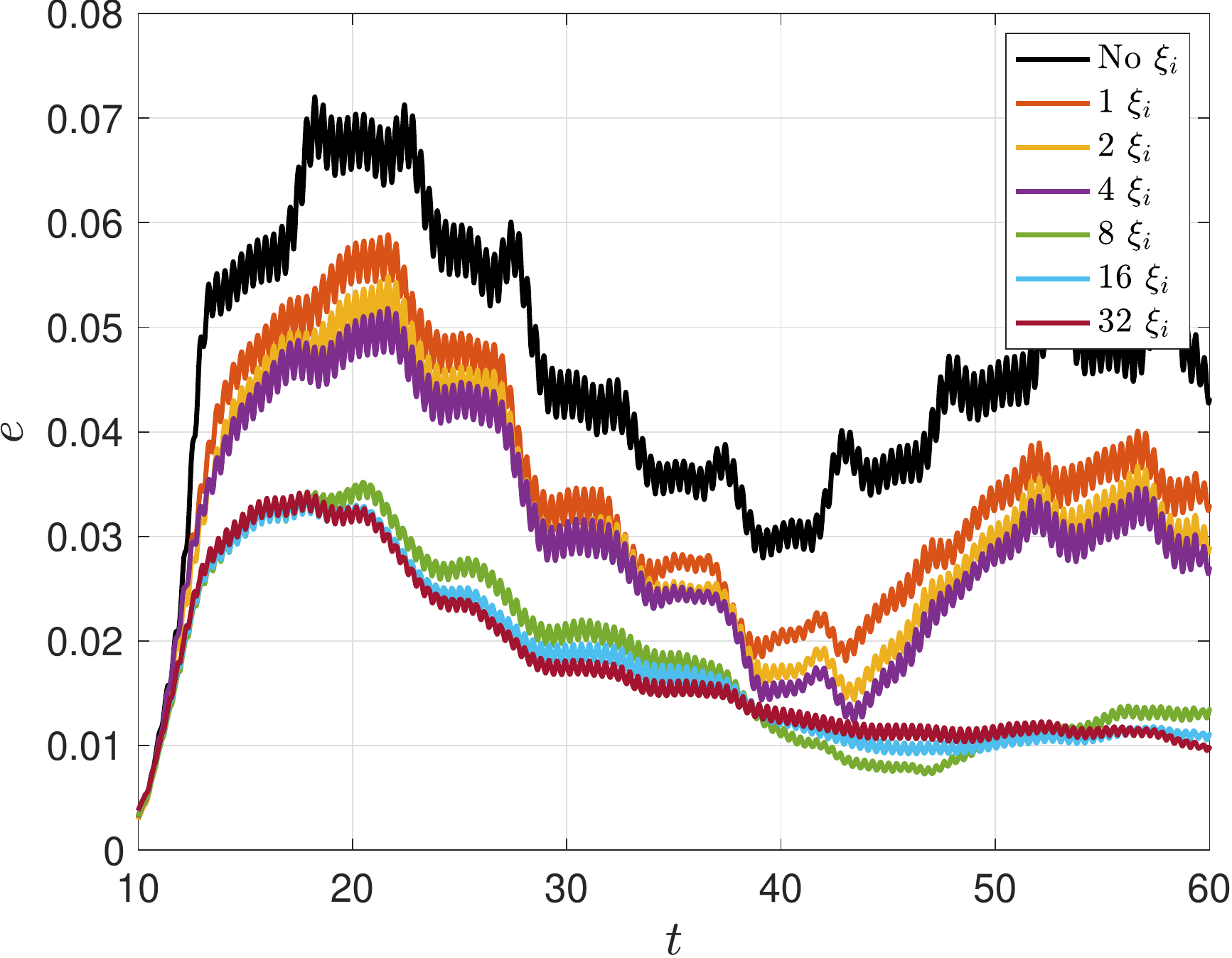} 
        \caption{~}
    \end{subfigure}
    \caption{Error \eqref{eq:ROMerror} on a grid consisting of 32 cells for an increasing number of EOFs using the FD method for perturbed IC 1 (a) and perturbed IC 2 (b).}
    \label{fig:error_reduction_perturbedICs}
\end{figure}

The measured errors for these initial conditions are given in figure \ref{fig:error_reduction_perturbedICs}. Application of the correction term leads to a decrease of the error, which becomes especially apparent when applying the correction to perturbed IC 1 while less so for perturbed IC 2. This behavior is to be expected, since the correction term was designed for one specific initial condition. However, the results presented in figure \ref{fig:error_reduction_perturbedICs} indicate that the measured temporal coefficients tolerate some level of approximation without a significant loss in error reduction. We note that a further reduction of the error may be achieved by constructing an estimation of the temporal coefficients \eqref{eq:temporalcoefficients_matrix} for the $\boldsymbol \xi_i$ and would further extend the applicability of the reduced-order correction terms. Examples of such methods have been suggested in literature, such as regarding the temporal coefficients as a stochastic process \cite{cotter2019numerically} or state-dependent subgrid forcing \cite{arnold2013stochastic}.

\subsection{Approximation of long-time averages} \label{subsec:ROM_statsteadystate}
Often, in practical situations, one does not wish to reproduce the instantaneous fields but rather long-time averages or statistics of the underlying fields. In this subsection, we study this problem by obtaining the EOFs and corresponding time series from a particular data set and subsequently applying the obtained forcing to the same flow with a different initial condition.

The EOFs and corresponding time series are computed for the second test case after the flow has reached a periodic regime due to the periodic forcing. The EOFs are measured for 10 time units, one period of the forcing, from the periodic state. In this regime, it has been verified that the EOFs are the same for each periodic measuring interval, as expected. Therefore, the change in the  initial condition only affects the temporal coefficients coefficients.

\begin{figure}[h!]
    \centering
    \begin{subfigure}{0.45\textwidth}
        \centering
        \includegraphics[width=\textwidth]{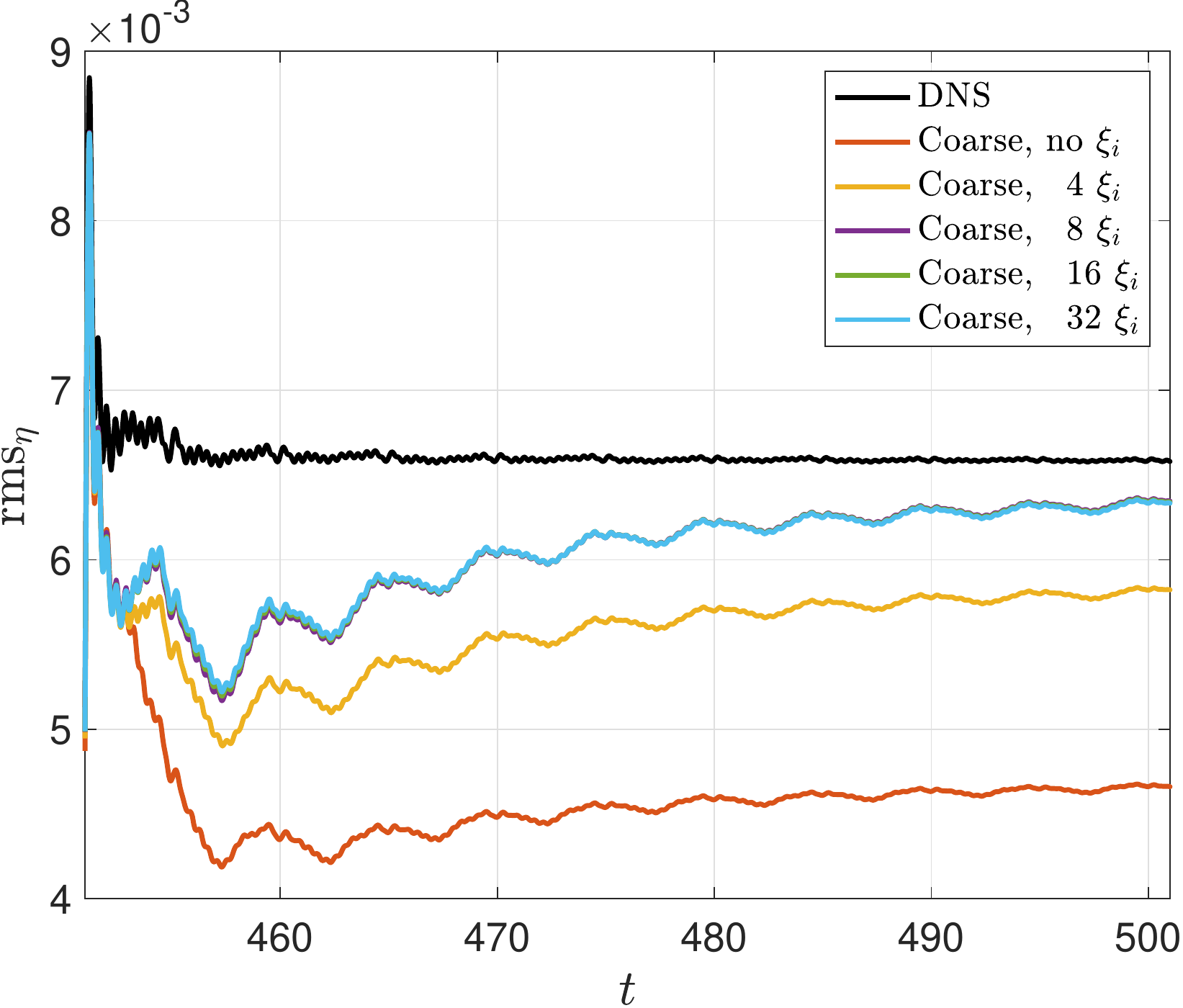}
        \caption{~}
    \end{subfigure}\hfill
    \begin{subfigure}{0.45\textwidth}
        \centering
        \includegraphics[width=\textwidth]{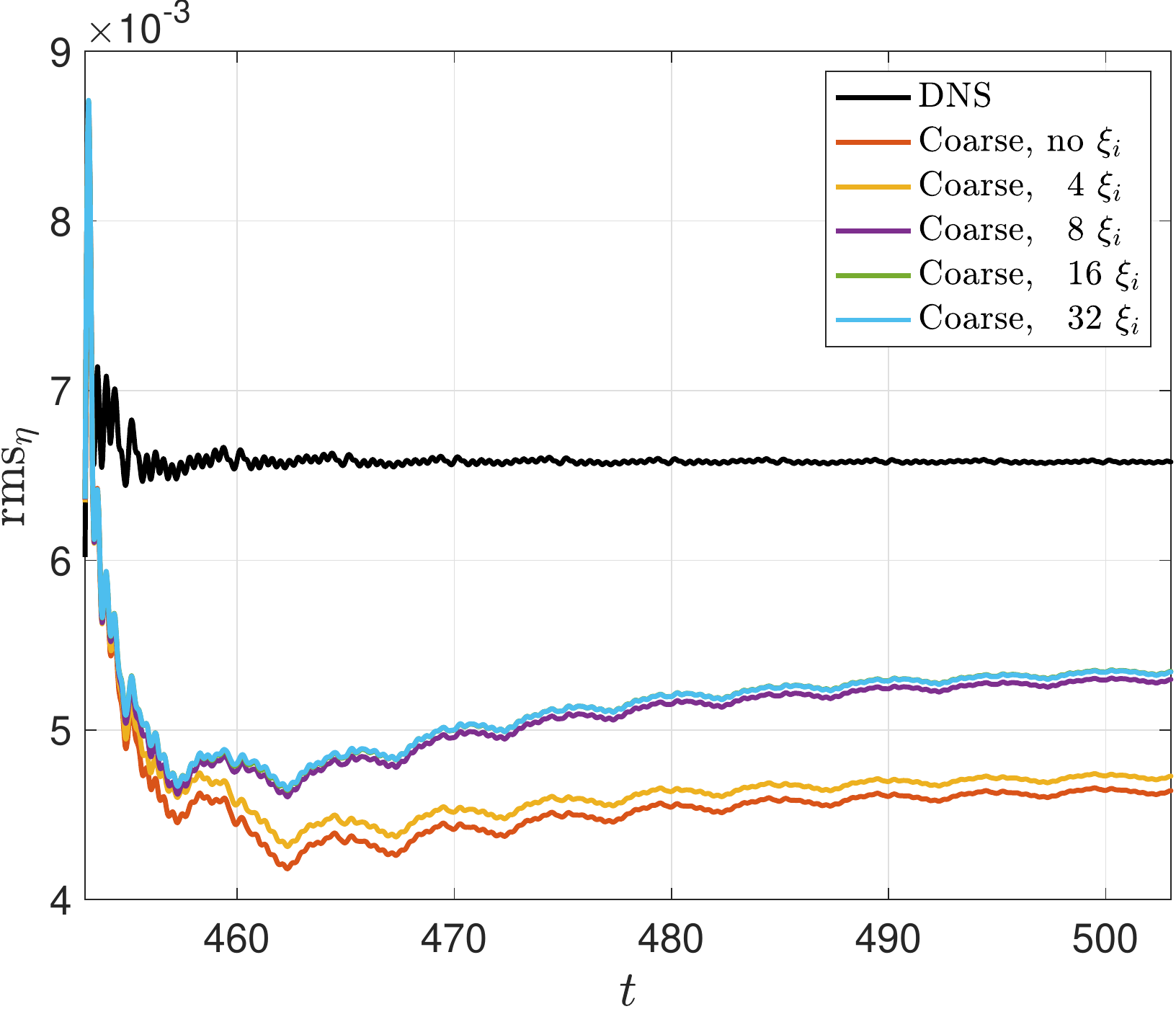} 
        \caption{~}
    \end{subfigure}
    \caption{Moving time-mean of the rms of $\eta$ for a different number of EOFs and for two initial conditions in the stationary regime. The initial conditions are phase-shifted by one time unit (a) and by three time units (b).}
    \label{fig:eta_variation}
\end{figure}

To study the ability of the measured corrections to approximate mean quantities of the flow, we compare the root mean square (rms) variation of the free surface height, \begin{equation}
\mathrm{rms}_{\eta}(t_k) = \left(\frac{1}{N}\sum_{j=1}^N\left[\eta(x_j,t_k) - \frac{1}{N}\sum_{i=1}^N \eta(x_i,t_k)\right]^2\right)^{1/2}. 
\end{equation}
We consider two initial conditions in the periodic regime. Compared to the measuring time, the first initial condition is phase-shifted by one time unit and the second initial condition is phase-shifted by three time units. Applying the measured corrections to these situations yields the rms of $\eta$ shown in figure \ref{fig:eta_variation}. It can be observed that including the correction terms leads to an improvement in the prediction for both cases, but the level of improvement depends on the chosen initial condition. This behavior is to be expected, since the correction terms are tailored for one specific situation. 

Analogous to what was shown in section \ref{subsec:ROM_ICs}, a further reduction of the error is expected to take place when a model able to also account for the current state of the solution is applied to the temporal coefficients. 

\newpage
\section{Conclusions and outlook}\label{sec:conclusion}
In this paper, we have compared subgrid measurements of the difference between a highly resolved truth and a corresponding coarse representation obtained with a finite difference and a finite element method for the one-dimensional shallow water equations. This difference was used to obtain a reduced-order correction on coarse grids. Special attention was given to the definition of these measurements, such that subgrid features caused by numerical error could be account for. This error draws contributions from both an incomplete representation of the spatial derivatives as well as from inaccuracies with which details in the bathymetry are included. The measurements of coarse-grid correction were decomposed into empirical orthogonal functions (EOFs) subsequently used to define a high-fidelity reduced-order model.

The EOFs were found to reflect the associated error of the particular discretization. While the reduced order correction can be constructed such that any numerical errors can in principle be fully eliminated for any discretization, the actual characteristics of the corrections are highly specific to the adopted discretization approach. Convergence of the subgrid corrections towards zero was observed for both discretization methods and for each eigenfunction with grid refinement. Going from coarser to finer grids, less of the available data is required to capture a certain fraction of the variability of the subgrid measurements. This procedure was applied to a steady case and a periodically forced case, for a given bathymetry.

The developed reduced-order correction has been defined such that the DNS representation on coarser grids could be reconstructed exactly. 
Here, this implies that the fine-grid solution on the coarse grid locations is captured fully. Even using only a fraction of the available EOFs for each state variable yields a significant improvement over the coarse grid solution. This procedure also identifies the weakest point in each discretization, by showing where one can improve the most upon using more EOFs. 

The reduced-order corrections were applied to several situations that differ from the original data set that the model was trained on. It was observed that predictions of mean quantities were improved when including the correction term. The level of improvement depends on the number of EOFs used in the model and on the distance from the initial condition of the data set. In addition, sensitivity to initial conditions was further explored and it was found that the corrections tolerate some level of approximation. This result makes it clear that accurately predicting the time series of each of the EOFs in the correction term will lead to further error reduction.

The results presented in this paper may be used in future work regarding coarse-grid predictions of geophysical fluid flows. Of particular interest is the application of the reduced-order correction for complex models such as the (thermal) rotating shallow water equations which are characterized by a richer dynamics that that of the sloshing case for the shallow water equations. Here we have shown that using a subgrid model constructed by a suitable subset of the EOFs of the actual subgrid term yields effective error reductions of coarse-grid predictions. This held true also in the situation for which initial conditions were not too far from those used for generating the dataset. The latter observation hints to the relevance of the modelling of the temporal coefficients once provided with an EOF basis from data. Additionally, the numerical experiments presented in this paper can be performed using different numerical methods to gain better understanding of the behaviour of the EOFs on coarse grids for different numerical methods. This can in turn lead to better predictions of the behavior of EOFs when DNS is not available, or when different flow conditions are considered.


\section*{Acknowledgements}
The authors would like to thank Darryl Holm and James-Michael Leahy, both at the Department of Mathematics, Imperial College London, for the many inspiring and clarifying discussions we had in the context of the SPRESTO project, funded by the Dutch Science Foundation (NWO) in their TOP1 program. 

\bibliographystyle{siamplain}
\bibliography{OneFile}

\appendix
\section{Description of the compatible finite element method}
\label{Appendix:FEM}
We present the FE approach in a number of steps.
Given the divergence-conforming space
\begin{equation}
    H(\text{div}, \Omega) = \{\mathbf{v} \in \big(L^2(\Omega)\big)^d | \nabla \cdot \mathbf{v} \in L^2(\Omega) \},
\end{equation}
where $\Omega$ denotes the (periodic) domain and $d$ its dimension, the function spaces $\mathbb{V}_u$ for the velocity field and $\mathbb{V}_\eta$ for total depth field are set up to satisfy
\begin{displaymath}
    \xymatrix{
        H(\text{{\normalfont div}};\Omega)\ar[d]^{\pi^1} \ar[r]^{\nabla \cdot} & L^2(\Omega) \ar[d]^{\pi^2} \\
        \mathbb{V}_u(\Omega) \ar[r]^{\nabla \cdot}  & \mathbb{V}_\eta(\Omega)}
\end{displaymath}\\
for bounded projections $\pi^1$, $\pi^2$ such that the diagram commutes. In the one-dimensional case, the divergence reduces to the single derivative $\partial_x$, and a pair of compatible spaces for $u$ and $\eta$ is given, e.g., by
\begin{equation}
    \mathbb{V}_u = CG_k(\Omega), \hspace{1cm} \mathbb{V}_\eta = DG_{k-1}(\Omega),
\end{equation}
where $CG_k(\Omega)$ denotes the $k^{th}$ polynomial order continuous Galerkin space and $DG_{k-1}(\Omega)$ the $(k-1)^{th}$ polynomial order discontinuous Galerkin space.

The governing shallow water equations \eqref{eq:rsw} can now be discretized such that the divergence in the continuity equation is considered strongly, while the gradient in the momentum equation is imposed weakly, leading to the mixed formulation
\begingroup
\addtolength{\jot}{2mm}
\begin{align}
    &\left\langle w, u_t \right\rangle - \left\langle w_x, \frac{1}{2}u^2 + \frac{1}{\text{Fr}^2}(\eta - b) \right\rangle = 0 & \forall w \in \mathbb{V}_u, \label{cfem_u_eqn_no_upw}\\
    & \eta_t + F_x = 0, \label{cfem_eta_eqn_no_upw}
\end{align}
\endgroup
where $\langle . ,. \rangle$ denotes the $L^2$ inner product, and the flux $F$ in~(\ref{cfem_eta_eqn_no_upw}) is given by the $L^2$-projection of $\eta u$ into the velocity space, i.e.,
\begin{align}
    \langle w, F - \eta u \rangle = 0 && \forall w \in \mathbb{V}_u.
\end{align}
In this so-called compatible framework, the continuity equation is formulated in strong form, as the derivative in $x$ maps the flux $F$ into $\mathbb{V}_\eta$. Further, no surface integral is required for the spatial derivative's weak formulation in the momentum equation, since $w_x \in \mathbb{V}_\eta$ is well-defined everywhere. The above space discretization conserves mass locally as well as a discrete energy globally (for details, see e.g. \cite{mcrae2014energy}).
Finally, we also incorporate transport stabilization for $\eta$ without compromising on the latter two conservation properties, by modifying equations \eqref{cfem_u_eqn_no_upw} - \eqref{cfem_eta_eqn_no_upw} according to \cite{wimmer2020energy}
\begingroup
\addtolength{\jot}{2mm}
\begin{align}
    &\left\langle w, u_t \right\rangle + \langle P_x, w \rangle - \int_\Gamma [\![P]\!]\left\{\frac{w}{\eta}\right\}\tilde{\eta} = 0 & \forall w \in \mathbb{V}_u, \label{cfem_u_eqn}\\
    & \langle \phi, \eta_t \rangle - \langle \phi_x, F \rangle + \int_\Gamma [\![\phi]\!]\left\{\frac{F}{\eta}\right\}\tilde{\eta} \; dS = 0 &\forall \phi \in \mathbb{V}_\eta, \label{cfem_eta_eqn}
\end{align}
\endgroup
where in a similar fashion to $F$, $P$ is given by an $L^2$-projection of the form
\begin{align}
    \left\langle \phi, P - \left(\frac{1}{2}u^2 + \frac{1}{\text{Fr}^2}(\eta - b) \right)\right\rangle = 0 && \forall \phi \in \mathbb{V}_\eta.
\end{align}
The integrals are over all cell boundaries (which in 1D reduces to evaluations at single points), and $[\![ . ]\!]$ and $\{.\}$ denote difference and average values, respectively. Finally, $\tilde{\eta}$ denotes the upwind value along the given cell boundary. Note that in the adopted Runge-Kutta scheme, the projections $F$ and $P$ need to be evaluated separately before each evaluation of the dynamic contribution. In this paper, we consider the lowest polynomial order $k=1$ for the mixed compatible setup.
The scheme and varying resolution mesh hierarchies are implemented using the automated finite element toolkit Firedrake, see \cite{rathgeber2016firedrake, Mitchell2016}\footnote{For further details, visit http://firedrakeproject.org }, which in turn relies on PETSc, see \cite{balay2019petsc, balay1997efficient}.

\end{document}